\documentclass{rspublic}

\usepackage[dvips]{graphicx}
\usepackage{amsmath,color}

\newcommand\ba{\begin{eqnarray}}
\newcommand\ea{\end{eqnarray}}
\newcommand\bc{\begin{center}}
\newcommand\ec{\end{center}}

\def\bi{\begin{itemize}}
\def\ei{\end{itemize}}
\def\bn{\begin{enumerate}}
\def\en{\end{enumerate}}
\def\bmp{\begin{minipage}}
\def\emp{\end{minipage}}
\def\nn{\nonumber}

\newcommand\p{\partial}

\newcommand{\bfu}{\ensuremath{\mathbf{u}}}
\newcommand{\bfe}{\ensuremath{\mathbf{e}}}
\newcommand{\bfF}{\ensuremath{\mathbf{F}}}
\newcommand{\bfphi}{\ensuremath{\boldsymbol{\Phi}}}
\newcommand{\bfomega}{\ensuremath{\boldsymbol{\omega}}}
\newcommand{\curl}{\ensuremath{\nabla\!\times}}
\newcommand\D{\ensuremath{\mathcal{D}}}
\newcommand\RR{\ensuremath{Re}}

\def\epsilon{\varepsilon}
\def\e{\mathrm{e}}

\def\e{\ensuremath{\mathrm{e}}}
\def\i{\ensuremath{\mathrm{i}}}

\def\lp{\left(}
\def\rp{\right)}

\newcommand{\KE}{\ensuremath{E}}

\begin{document}
\title[Transition in a 1D shear flow model]
{Turbulent transition in a truncated 1D model for shear flow}

\author{J. H. P. Dawes and W. J. Giles}

\affiliation{Department of Mathematical Sciences, University of Bath,
Claverton Down, Bath BA2 7AY, UK}

\label{firstpage}
\maketitle

\begin{abstract}{fluid flow; turbulence; dynamical systems}
We present a reduced model for the transition to turbulence
in shear flow that is simple enough to admit a thorough numerical
investigation while allowing spatio-temporal dynamics that
are substantially more complex than those allowed in previous
modal truncations.

Our model allows a comparison of
the dynamics resulting from initial perturbations
that are localised in the spanwise direction with
those resulting from sinusoidal perturbations.
For spanwise-localised initial conditions the subcritical
transition to a `turbulent' state (i) takes place more
abruptly, with a boundary between laminar
and `turbulent' flow that is appears to be much less `structured'
and (ii) results in a spatiotemporally chaotic
regime within which the lifetimes of spatiotemporally
complicated transients are longer, and are
even more sensitive to initial conditions.

The minimum initial energy $E_0$ required
for a spanwise-localised initial perturbation
to excite a chaotic transient has a power-law scaling with Reynolds
number $E_0 \sim Re^p$ with $p \approx -4.3$. The
exponent $p$ depends only weakly on the width of the localised
perturbation and is lower than that commonly observed in
previous low-dimensional models where typically $p \approx -2$.

The distributions of lifetimes of chaotic
transients at fixed Reynolds number are
found to be consistent with exponential distributions.
\end{abstract}

%PACS Codes: 47.20 (bifurcation theory) ; 47.54 (pattern formation).

%\begin{figure}
%\begin{center}
%\includegraphics[angle=270,width=11.0cm]{}
%\caption{}
%\label{fig:linear_pic}
%\end{center}
%\end{figure}

%%%%%%%%%%%%%%%%%%%%%%%%%%%%%%%%%%%%%%%%%%%%%%%%%%%%%%%%%%%%%%%%%%%%%%%%%%%%%%%%%%%%%%%%%%

\section{Introduction}

The transition from laminar to turbulent states has
been a central problem in fluid mechanics for many decades.
Since the 1960s, promising lines of attack have been
opened up through the use of ideas from dynamical systems theory
coupled to the thorough investigation of reduced versions of
the Navier--Stokes equations. Perhaps the best-known example
of such a reduction is the derivation of the Lorenz equations
as a model of the onset of thermal convection in a layer of
fluid heated from below (Lorenz 1963).
Although the set of three nonlinear
ODEs that comprise the Lorenz 1963 model displays
a wealth of interesting dynamical behaviour, little of this
is relevant to the original fluid-mechanical problem. However,
similar approaches yield convincing agreement over wide
ranges of parameter values in other fluid-mechanical
situations, for example thermal convection in the presence
of a magnetic field, and the onset of Taylor vortices
in the flow between rotating coaxial concentric cylinders.
In situations such as these the flow undergoes a series
of bifurcations before becoming chaotic, in a sense that
can be given a clear meaning in terms of the behaviour of the
reduced model comprising nonlinear ODEs.

In contrast, the transition to turbulence in shear flows
appears not to proceed through a sequence of bifurcations, but
to be linked to the appearance, at a critical
Reynolds number $Re_c$, of a chaotic saddle in phase space:
a complicated collection of unstable equilibria and time-periodic orbits
which results in ever longer transients before the flow relaxes
to a purely laminar state.
For a review, and substantial numbers
of references to the literature (although with
an emphasis on pipe flow), see Kerswell (2005).
The significance of $Re_c$ is that for $Re<Re_c$ the laminar state is
a global attractor and trajectories appear to evolve
rapidly towards it, while for $Re>Re_c$ other (usually unstable)
invariant sets exist in phase space. These new invariant sets cause
increasingly long transient excursions to take place 
before relaminarisation occurs, for initial conditions that are
sufficiently far from the laminar profile. Small perturbations
to the laminar state still decay rapidly towards it, and there
appears to be a distinct boundary separating the behaviour of
trajectories into those that relax rapidly and those which undergo
long transient excursions. This laminar-turbulent
boundary is sometimes referred to as
the `edge of chaos'.
Despite these differences in phenomenology and dynamics,
reduced models constructed
along the same lines as the Lorenz model have provided
substantial insight into both the physical origin of
this transition to self-sustaining complicated motion, and
the mathematical organisation of equilibria and periodic orbits
inside the chaotic saddle.

The study by Waleffe (1997) provides the direct inspiration
for the present work. Waleffe showed that a low-order
reduced model could be constructed that elucidated the
different elements of a self-sustaining process (SSP) that
allowed sufficiently large deviations from the laminar
flow profile to persist indefinitely. The SSP can be
briefly described as follows. Weak streamwise vortices (i.e.
vortical rolls whose axes are aligned with the primary flow
direction) distort the streamwise velocity profile by
moving high and low-speed fluid around. This distortion
generates streamwise streaks of fluid that are moving faster
and slower than the fluid around them. The streaks are unstable
to modes which create vortical eddies
oriented in the wall-normal direction, orthogonal to the
streamwise vortices, and the resulting three-dimensional
re-organisation of the flow reinforces the streamwise
vortices.

Waleffe's model simplified the Navier--Stokes equations
first by considering, not plane Couette flow between rigid
boundaries, but a modified problem in which the boundaries
are stress-free and the laminar profile is sinusoidal, sustained
by an artificially-applied pressure term. It appears that
the physics of the SSP is rather insensitive to these modifications.
The second set of simplifications concern the Galerkin expansion
of the velocity field in all three directions: streamwise ($x$),
wall normal ($y$) and spanwise ($z$). Thus Waleffe considered
a model comprising 8 of the lowest-wavenumber Fourier
modes in this Galerkin expansion. These
8 modes were chosen in order to capture the central
elements of the SSP, and to be self-consistent in the sense
that the nonlinear (quadratic)
interactions between these 8 modes preserve
energy just as the full advective nonlinearity in the
Navier--Stokes equations does.
Having projected out the spatial dependence of the
dynamics onto these modes, the problem reduces to a far
simpler set of
ordinary differential equations (ODEs) for the
time-dependent mode amplitudes.
Subsequent work, in particular by
Eckhardt \& Mersmann (1999) and 
Moehlis, Faisst and Eckhardt (2004, 2005)
extended Waleffe's model
to include an additional physical insight: that the basic
laminar profile of the shear flow will itself
be modified by the nonlinear interactions with streamwise vortices
and streaks. Moehlis et al developed a 9-mode ODE model that
was amenable to investigation in substantial detail. In particular
they discussed the lifetimes of perturbations as the Reynolds
number increased, locating the onset of complicated dynamics,
and they also discussed the probabilistic
distribution of lifetimes at fixed Reynolds number
from randomised initial conditions of equal energy.

It is clear that the assumptions made by these authors
concerning spatial periodicity seems much easier to argue
for in the streamwise and wall-normal directions than in
the spanwise direction; indeed recent work by
Schneider and co-workers (Schneider et al 2009; 2010a; 2010b)
(see also Duguet, Schlatter and Henningson, 2009)
has concentrated on understanding
the formation of structures which are \textit{spatially localized\/}
in $z$: very far from periodic in this direction.

With this in mind we propose in this paper an extension of
Waleffe's model which is a reduction of the Navier--Stokes
equations to a collection of PDEs in $z$ and $t$: we adopt
the Fourier mode truncations that Waleffe used in order
to remove the dependence on the streamwise $x$ and wall-normal $y$
coordinates since numerical work shows that, at least for
some of these equilibrium and periodic orbit states,
the flow structure in these coordinates can be
well-approximated by a small number of Fourier modes.

By retaining full dependence on the spanwise coordinate $z$
we admit both the spatially-periodic solutions of Waleffe
and the formation of localized states. In addition,
a wealth of spatio-temporal complexity is allowed.
Our study is very similar in spirit to work by Manneville and
co-authors (Manneville \& Locher 2000;
Manneville 2004; Lagha \& Manneville 2007) who preserve full
resolution in two directions (streamwise and spanwise) and use
Galerkin truncation only in the third (`wall-normal') direction. This
enables these authors to consider the dynamics around turbulent spots that
are localised in both the streamwise and spanwise directions, at
the cost of more intensive numerical computations. Their work
is therefore complementary to that presented here.
We remark also that reduced models, of different kinds,
have been used both in pipe
flow (Willis \& Kerswell 2009) and in understanding the
formation of turbulent-laminar bands in plane Couette flow
(Barkley \& Tuckerman 2007).

The structure of the paper is as follows.
In section~\ref{sec:model} we discuss the derivation
of our PDE extension of Waleffe's ODE model. In section~\ref{sec:results}
we present the results of our numerical investigations
into the transition to turbulence described by the model.
We conclude in section~\ref{sec:conc}.

%%%%%%%%%%%%%%%%%%%%%%%%%%%%%%%%%%%%%%%%%%%%%%%%%%%%%%%%%%%%%%%%%%%%%%%%
\section{Derivation of the PDE model}
\label{sec:model}

In this section we define sinusoidal shear flow and summarise
the Galerkin truncation that we use to derive our simplified model
for turbulent transition.

\begin{figure}[!h]
\bc
\includegraphics[width=8.0cm]{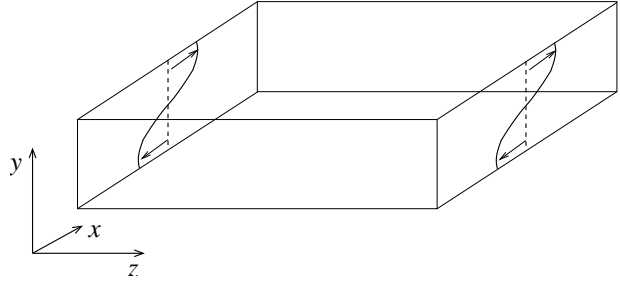}
\caption{Geometry of the domain, illustrating the basic
sinusoidal shear profile $\bfu=\sqrt{2} \sin \beta y \bfe_x$
which is sustained by the applied body force term.}
\label{fig:setup}
\ec
\end{figure}

Following Waleffe (1997) and Moehlis et al (2004),
we use the usual Cartesian
coordinate conventions that $x$ is the downstream
direction (`streamwise'), $y$ is the direction of the shear
gradient, i.e.
normal to the sidewalls and $z$ is the spanwise direction. We
write the Navier--Stokes equations for incompressible flow
in the nondimensionalised form
\ba
\frac{\p \bfu}{\p t} + \bfu \cdot \nabla \bfu = -\nabla p
+ \bfF(y) + \frac{1}{\RR} \nabla^2 \bfu, \label{eqn:ns}
\ea
where we have scaled lengths by $h/2$ where $h$ is the width of the
channel, velocities by $U_0$ the velocity of the laminar profile at a distance
$h/4$ from the upper boundary and pressure by $U_0^2 \rho$ where
$\rho$ is the density of the fluid. The Reynolds number $\RR$ is therefore
given by $\RR=U_0 h/(2\nu)$, $\nu$ being the kinematic viscosity.
The evolution of~(\ref{eqn:ns}) is subject to the usual incompressibility
condition $\nabla \cdot \bfu=0$ and the conditions
for impermeable and stress-free upper and lower boundaries
\ba
u_y =0, \qquad \mathrm{and} \qquad 
\frac{\p u_x}{\p y} = \frac{\p u_z}{\p y} =0 \qquad \mathrm{at}
\qquad y=\pm 1. \label{eqn:bcs}
\ea
The flow is assumed to be periodic in the $x$ and $z$ directions,
with periodicities $L_x$ and $L_z$ respectively.
We take the body force term $\bfF(y)$ to be
\ba
\bfF(y) & = & \frac{\sqrt{2}\beta^2}{\RR} \sin \beta y \, \bfe_x,
\label{eqn:bodyforce}
\ea
where $\bfe_x$ is the unit vector in $x$ direction and it is
convenient to define the constant $\beta=\pi/2$. The laminar
profile is then the steady solution
\ba
\bfu & = & \sqrt{2} \sin \beta y \bfe_x, \label{eqn:laminar}
\ea
of~(\ref{eqn:ns}) - (\ref{eqn:bodyforce}), as shown in figure~\ref{fig:setup}.
We note that although the `sinusoidal shear flow'
profile~(\ref{eqn:laminar}) has an inflection point, the
flow is linearly stable for all $\RR$ (Drazin \& Reid 1981).

We now turn to our solution ansatz. We write the velocity
field in the form
\ba
\bfu & = & \bfu_M + \curl \bfphi_T + \curl \curl \bfphi_P, \label{eqn:u}
\ea
where the subscripts denote mean, toroidal and poloidal components
which are, respectively, expressed as sums of the first few
Fourier modes in each case:
\ba
\bfu_M & = &  \left( A_1 \sin \beta y + A_2 \right) \bfe_x, \nn \\
\bfphi_T & = & A_3 \cos \beta y \, \bfe_x 
\nn \\ & & 
+ \left( 
A_4 \sin \alpha x - A_5 \cos \alpha x \sin \beta y
-A_6 \cos \alpha x + A_7 \sin \alpha x \sin \beta y
\right) \bfe_y, \nn \\
\bfphi_P & = & A_8 \cos \alpha x \cos \beta y \, \bfe_y. \nn
\ea
We define the streamwise and spanwise wavenumbers
$\alpha = 2\pi/L_x$ and $\gamma=2\pi/L_z$ for notational convenience.
The amplitudes $A_1,\ldots, A_8$ are functions of $z$ and $t$
whose evolution can be obtained by substituting~(\ref{eqn:u})
into~(\ref{eqn:ns}). Note that the form of the ansatz
implies that incompressibility and the boundary
conditions~(\ref{eqn:bcs}) are automatically satisfied.

The amplitudes $A_1,\ldots,A_8$ correspond exactly, in terms
of their Fourier dependence in $x$ and $y$, to
the modes selected by Waleffe (1997). For reference, 
table~\ref{table:notation} summarises the correspondence.
\begin{table}
\caption{Comparison of notation.}
\begin{tabular}{lcccccccc}
%\hline
\hline
\\
This paper     & $A_1$ & $A_2$ & $A_3$ & $A_4$ & $A_5$ & $A_6$ & $A_7$ & $A_8$ \\
Waleffe (1997) &   $M$ &   $U$ &   $V$ &   $A$ &   $C$ &   $B$ &   $D$ &   $E$ \\
& const & $\cos \gamma z$ & $\sin \gamma z$ & const & const & $\cos \gamma z$
& $\cos \gamma z$ & $\sin \gamma z$
\\
\hline
\end{tabular}
\label{table:notation}
\end{table}

We now briefly describe the role that each mode plays in the
dynamics. $A_1$ is the amplitude of the sinusoidal shear profile;
the laminar state corresponds to $A_1=\sqrt{2}$, $A_2=\cdots=A_8=0$.
$A_2$ describes variations in $z$ of the streamwise velocity, i.e.
the formation of streamwise streaks. $A_3$ describes the formation
of $x$-independent streamwise vortices that redistribute the shear profile.
Modes $A_4,\ldots,A_7$ describe, as in Waleffe (1997) the development
of $x$-dependent distortions of the streaks, and in particular the linear
instability of the $x$-independent streaks described by $A_2$. These
modes have no velocity component in the vertical (i.e. $y$) direction.
$A_8$ describes `oblique rolls' and, in contrast to modes $A_4,\ldots,
A_7$, has a non-zero vertical velocity component but no vertical
vorticity component.

To derive evolution equations for the modes $A_1,\ldots,A_8$ we
use Fourier orthogonality in the $x$ and $y$ directions combined with
projections onto individual components of either the velocity field $\bfu$,
the vorticity field $\bfomega=\curl\bfu$ or (for $A_8$)
the curl of the vorticity field. For later convenience
we introduce the differential
operators $\D_\alpha^2$, $\D_\beta^2$ and $\D_{\alpha\beta}^2$
which correspond to the action of $-\nabla^2$ on different Fourier modes:
\ba
\D_\alpha^2        %& 
\equiv
%&
 \alpha^2 - \p_{zz}, \quad
%\nn \\
\D_\beta^2         %&
 \equiv 
%&
 \beta^2  - \p_{zz}, \quad
%\nn \\
\D_{\alpha\beta}^2 %& 
\equiv 
%&
 \alpha^2 + \beta^2 - \p_{zz}. \nn
\ea
We denote $\p A_j/\p z$ by $A_j'$ for $j=1,\ldots, 8$.
Considering first the $\bfe_x$ component of $\bfu$ in~(\ref{eqn:ns})
we obtain the following PDEs in $z$ and $t$ for $A_1$ 
and $A_2$:
\ba
\lp \p_t + \frac{1}{\RR} \D_\beta^2 \rp A_1 & = & 
-\beta A_2'A_3
+ \frac{\alpha}{2}(A_4'' A_5 - A_4 A_5'')
+ \frac{\alpha}{2}(A_6 A_7'' - A_6'' A_7)
\nn \\ & &
+\frac{\beta}{2}(A_6'' A_8' - \alpha^2 A_6 A_8')
+ \frac{\sqrt{2}\beta^2}{\RR},
\label{eqn:a1} \\ \nn \\
\lp \p_t - \frac{1}{\RR} \p_{zz} \rp A_2 & = &
-\frac{\beta}{2} (A_1 A_3)'
+ \frac{\alpha}{2}(A_4'' A_6 - A_4 A_6'') 
-\frac{\alpha}{4}(A_5'' A_7 - A_5 A_7'') \nn \\
& & -\frac{\alpha^2\beta}{4}(A_5 A_8)'
+ \frac{\beta}{4}(A_5' A_8')',
\label{eqn:a2}
\ea
Now we turn to the vorticity $\bfomega$. For $A_3$, $A_4$
and $A_6$ we find that only one component of $\bfomega$
contains a contribution from each of these; taking
the curl of~(\ref{eqn:u}) the terms involving
$A_3$, $A_4$ and $A_6$ are
\ba
\bfomega & = & \D_\beta^2 A_3 \cos \beta y \, \bfe_x
+ \left(
\D_\alpha^2 A_4 \sin \alpha x - \D_\alpha^2 A_6 \cos \alpha x
\right) \bfe_y. \nn
\ea
Therefore it is straightforward to consider the $x$ and $y$ components
of the vorticity equation obtained by applying
the operators $\bfe_x \cdot \curl$ and $\bfe_y \cdot \curl$
to~(\ref{eqn:ns}) in order
to obtain evolution equations for $A_3$, $A_4$ and $A_6$:
\ba
\lp \p_t + \frac{1}{\RR} \D_\beta^2 \rp \D_\beta^2 A_3 & = & 
-\alpha^2\beta(A_4 A_7)' -\alpha^2\beta(A_{5} A_{6})'
+\frac{\alpha^3}{2}(A_4 A_8)'' \nn \\ & & 
- \frac{\alpha}{2}(A_4 A_8'')''
+\alpha \beta^2 A_4' A_8' + \frac{\alpha^3\beta^2}{2}A_4 A_8
\nn \\ & & 
+ \frac{\alpha\beta^2}{2}A_4 A_8'',
\label{eqn:a3}
\ea
\ba
\lp \p_t + \frac{1}{\RR} \D_\alpha^2 \rp \D_\alpha^2 A_4 & = & 
\frac{\alpha}{2}(A_1 A_5'' - A_1'' A_5) - \frac{\alpha^3}{2}A_1 A_5
+\alpha (A_2 A_6'' - A_2'' A_6) \nn \\ & & 
- \alpha^3 A_2 A_6
+\frac{\beta}{2}(A_3 A_7')'' - \frac{\alpha^2\beta}{2}A_3 A_7'
-\alpha^2\beta A_3' A_7 
\nn \\ & &
-\frac{\alpha^3\beta^2}{2}A_3 A_8
+ \frac{\alpha\beta^2}{2}(A_3 A_8'' - A_3'' A_8),
\label{eqn:a4}
\ea
\ba
\lp \p_t + \frac{1}{\RR} \D_\alpha^2 \rp \D_\alpha^2 A_6 & = &
\frac{\alpha^3}{2}A_1 A_7 - \frac{\alpha}{2}(A_1 A_7'' - A_1'' A_7)
+\alpha^2 \beta A_1' A_8 
\nn \\ & & 
+ \frac{\alpha^2 \beta}{2}A_1 A_8'
- \frac{\beta}{2}(A_1 A_8')''
+\alpha^3 A_2 A_4 \nn \\
& & -\alpha(A_2 A_4'' - A_2'' A_4)
+\frac{\beta}{2} (A_3 A_5')''
- \alpha^2 \beta A_3' A_5
\nn \\ & & 
-\frac{\alpha^2\beta}{2} A_3 A_5',
\label{eqn:a6}
\ea

Evolution equations for $A_5$, $A_7$ and $A_8$ are
derived similarly, using (for $A_5$ and $A_7$) the projection
operator $\bfe_y \cdot \curl$
and (for $A_8$) the projection operator
$\bfe_y \cdot \curl \curl$
applied to~(\ref{eqn:ns}). The resulting evolution equations are
\ba
\lp \p_t + \frac{1}{\RR} \D_{\alpha\beta}^2 \rp \D_\alpha^2 A_5 & = & 
\alpha^3 A_1 A_4 - \alpha (A_1 A_4'' - A_1'' A_4)
+\alpha^3 A_2 A_7 \nn \\ & & 
- \alpha (A_2 A_7'' - A_2'' A_7)
-\beta (A_2' A_8')' 
\nn \\ & & 
+ \alpha^2 \beta A_2' A_8
-\alpha^2\beta (A_3 A_6)' + \beta (A_3 A_6'')',
\label{eqn:a5} \\
\lp \p_t + \frac{1}{\RR} \D_{\alpha\beta}^2 \rp \D_\alpha^2 A_7 & = & 
-\alpha^3 A_1 A_6 + \alpha(A_1 A_6'' - A_1'' A_6)
-\alpha^3 A_2 A_5 \nn \\ & & 
+ \alpha(A_2 A_5'' - A_2'' A_5)
+\beta (A_3 A_4'')' 
\nn \\ & &
- \alpha^2 \beta (A_3 A_4)',
\label{eqn:a7} \\
\lp \p_t + \frac{1}{\RR} \D_{\alpha\beta}^2 \rp \D_{\alpha\beta}^2
\D_\alpha^2 A_8 & = &
-2\alpha^2 \beta (A_1' A_6 + A_2' A_5) -2\alpha \beta^2 A_3' A_4'
\nn \\ & & 
+\alpha (A_3'' A_4)'' 
- \alpha(\alpha^2+\beta^2)A_3'' A_4.
\label{eqn:a8}
\ea
Equations~(\ref{eqn:a1}) - (\ref{eqn:a8}) are a closed set
of nonlinear
PDEs in $z$ and $t$ which form a truncated model of the
dynamics of sinusoidal shear flow.
Crucially these equations satisfy two consistency checks:
the nonlinear terms in~(\ref{eqn:a1}) - (\ref{eqn:a8})
conserve energy and so reflect the conservative nature of the full
$\bfu \cdot \nabla \bfu$ nonlinearity in~(\ref{eqn:ns}).
Secondly
this model reduces to the model of
Waleffe (1997) in the special case in which
the amplitudes $A_1,\ldots,A_8$ are
taken to be periodic in the $z$-direction, after applying
appropriate projection onto orthogonal Fourier modes in $z$.
We discuss each of these important issues in more detail
in the following subsections.

%%%%%%%%%%%%%%%%%%%%%%%%%%%%%%%%%%%%%%%%%%%%%%%%%%%%%%%%%%%%%%%%%
\subsection{Nonlinear terms and energy conservation}

In this subsection we show that the nonlinear terms in
~(\ref{eqn:a1}) - (\ref{eqn:a8}) do not contribute to the energy
budget for the dynamics. This is a physically crucial property
in constructing any reasonable reduced model for shear
flows: the only energy source term
must be the body force $\bfF(y)$ that drives the laminar flow, and
the only source of dissipation must be (linear) viscous diffusion.

We define the (dimensionless) total kinetic energy of the flow to be
\ba
\KE & = & \frac{1}{2} \int_{\Omega} \bfu \cdot \bfu \ dx\, dy\, dz, \label{eqn:ke}
\ea
where $\Omega=[0, \pi/\alpha] \times [-1,1] \times
[0, L_z]$ is (one half of) the domain (in $x,y,z$ coordinates)
occupied by the fluid. We substitute our ansatz~(\ref{eqn:u})
into~(\ref{eqn:ke}) and carry out the $x$ and $y$ integrals, noting
that Fourier orthogonality enables us to remove all cross-terms
except those involving $A_7$ and $A_8$ since they have the same
Fourier dependencies; their contributions to $\bfu$ are
\ba
\bfu_7 = \left(
\begin{array}{c}
-A_7' \sin \alpha x \sin \beta y \\
0 \\
\alpha A_7 \cos \alpha x \sin \beta y 
\end{array}
\right),
\qquad \mathrm{and} \qquad
\bfu_8 = \left(
\begin{array}{c}
-\alpha \beta A_8 \sin \alpha x \sin \beta y \\
(\alpha^2 A_8 - A_8'') \cos \alpha x \cos \beta y \\
-\beta A_8' \cos \alpha x \sin \beta y 
\end{array}
\right). \nn
\ea
Hence the kinetic energy $\KE$ is given by
\ba
\KE & = & \frac{\pi}{2\alpha} \int_{0}^{L_z}
A_1^2 + 2A_2^2 + (A_3')^2 + \beta^2 A_3^2
+ (A_4')^2 + \alpha^2 A_4^2 \nn \\
& & 
+ \frac{1}{2} (A_5')^2 + \frac{\alpha^2}{2}A_5^2
+ (A_6')^2 + \alpha^2 A_6^2
+ \frac{1}{2}\left( \alpha \beta A_8 - A_7'\right)^2 \nn \\ & &
+ \frac{1}{2}\left( \alpha^2 A_8-A_8'' \right)^2
+ \frac{1}{2}\left( \alpha A_7 - \beta A_8'\right)^2
\, dz, \nn \\
& \equiv & \frac{\pi}{2\alpha} \int_{0}^{L_z}
\tilde E(z,t) \, dz, \label{eqn:etilde}
\ea
which defines the `local' energy quantity $\tilde E(z,t)$.
After integrating several terms by parts, and noting that
the boundary contributions vanish (since we
use a periodic boundary condition in $z$),
and also after some manipulation of the cross-terms involving
$A_7$ and $A_8$, we obtain the expression
\ba
\KE & = & \frac{\pi}{2\alpha} \int_{0}^{L_z}
A_1^2 + 2A_2^2 + A_3 \D_\beta^2 A_3 + A_4 \D_\alpha^2 A_4
+\frac{1}{2}A_5 \D_\alpha^2 A_5 \nn \\
& & + A_6 \D_\alpha^2 A_6
+\frac{1}{2} A_7 \D_\alpha^2 A_7 + \frac{1}{2} A_8 \D_{\alpha\beta}^2
\D_\alpha^2 A_8
\, dz. \nn
\ea
We can now compute the time evolution of the kinetic energy
by differentiating with respect to time. After carrying out further
integrations by parts we find
\ba
\frac{d \KE}{dt} & = & \frac{\pi}{2\alpha} \int_{0}^{L_z}
2 A_1 \dot A_1 + 4 A_2 \dot A_2 + 2 A_3 \D_\beta^2 \dot A_3
+2 A_4 \D_\alpha^2 \dot A_4 \nn \\ 
& & + A_5 \D_\alpha^2 \dot A_5
+2 A_6 \D_\alpha^2 \dot A_6 + A_7 \D_\alpha^2 \dot A_7
+ A_8 \D_{\alpha\beta}^2 \D_\alpha^2 \dot A_8
\, dz. \label{eqn:ke2}
\ea
We now substitute for the time derivatives
using the PDEs~(\ref{eqn:a1}) - (\ref{eqn:a8}). The interest
in pursuing this calculation is that at this
stage we find that the conservative nature of the quadratic
nonlinearities in~(\ref{eqn:a1}) - (\ref{eqn:a8}) becomes
apparent since all the cubic terms cancel (after appropriate
integrations by parts).
We are then left with a single linear source term
and a collection of quadratic dissipation terms:
\ba
\frac{d\KE}{dt} & = & \frac{\pi \beta^2 \sqrt{2}}{\alpha \RR}
\int_{0}^{L_z} A_1 \, dz
- \frac{\pi}{2\alpha \RR} \int_{0}^{L_z}
2\left( \beta^2 A_1^2 + (A_1')^2 \right)
+ 4(A_2')^2 \nn \\ 
& & + 2(\D_\beta^2 A_3)^2 +2 (\D_\alpha^2 A_4)^2 + (\D_\alpha^2 A_5)^2
+\beta^2 \left( \alpha^2 A_5^2 + (A_5')^2\right)
\nn \\
& & +2(\D_\alpha^2 A_6)^2 + (\D_\alpha^2 A_7)^2
+\beta^2 \left( \alpha^2 A_7^2 + (A_7')^2 \right)
\nn \\ 
& & +\alpha^2 (\D_{\alpha\beta}^2 A_8)^2 + (\D_{\alpha\beta}^2 A_8')^2
\, dz. \nn
\ea
The derivative of this equation for the evolution of the total kinetic
energy, which describes the balance between the driving provided
by the body force term $\bfF(y)$ and viscous dissipation, makes it
clear that the nonlinear terms in~(\ref{eqn:a1}) - (\ref{eqn:a8})
conserve energy. This justifies the self-consistency of the
selection of just the eight
modes $A_1,\ldots,A_8$ in the reduced model: by including
exactly these modes, no
unphysical effects are introduced into the energy evolution.

%%%%%%%%%%%%%%%%%%%%%%%%%%%%%%%%%%%%%%%%%%%%%%%%%%%%%%%%%%%%%%%%%%%%%%
\subsection{Relation with the model of Waleffe (1997)}

Waleffe (1997) considered a far simpler modal truncation in which
each mode contains only a single Fourier mode in $z$. This introduces
an additional parameter: the wavenumber $\gamma$ in the $z$-direction,
but the model comprises ODEs rather than PDEs
for the eight amplitudes and is
therefore much more amenable to analysis.
The derivation of such a modal truncation is slightly simpler than
that described above since one can use Fourier orthogonality directly
in the $z$-direction as well as in $x$ and $y$. In fact, the
reduced model~(\ref{eqn:a1}) - (\ref{eqn:a8}) reduces exactly
(after rescalings of the amplitude variables) to that discussed
by Waleffe if one inserts the corresponding trigonometric dependencies
and then projects out unwanted modes that arise from some of the
nonlinear terms. This projection step is self-consistent
in the sense that in both~(\ref{eqn:a1}) - (\ref{eqn:a8}) and the
ODEs derived by Waleffe the nonlinear terms conserve energy.
Table~\ref{table:notation} lists the Fourier
dependence in the $z$-direction that Waleffe (1997) assumed for each
mode.

Waleffe observed that his ODE model had a number of 
deficiencies, for example the streaks (described by $A_2$) were
unstable to $x$-dependent perturbations, with even and
odd symmetry in $y$, only if the wavenumber $\gamma$
was sufficiently large: $\gamma^2>\alpha^2$ 
and $\gamma^2>\alpha^2+\beta^2$, respectively, to be precise.

More seriously, he discusses
the inadequacy of his ODE model in capturing the interaction
between the mean shear ($M$, or equivalently $A_1$)
and the $x$-dependent modes ($A$, $C$, $B$ and $D$,
or equivalently $A_4,\ldots,A_7$) that arise from
consideration of advection by the mean shear. We observe that in
~(\ref{eqn:a1}) - (\ref{eqn:a8}) these quadratic interactions take
far more complex forms that in many cases vanish identically
when the simple Fourier mode dependencies in $z$ that are listed
in table~\ref{table:notation} are imposed. In particular
the mean shear $A_1$ is influenced by new
combinations of $x$-dependent
modes which do not arise in the ODE model because the
expressions $A_6 A_7'' - A_6'' A_7$ and $A_4'' A_5 - A_4 A_5''$ which
are present in~(\ref{eqn:a1}) vanish
identically in the ODE reduction. Terms with a structure identical
to this (i.e. of the form $A_n A_m''-A_n'' A_m$)
appear in several of the the other amplitude equations.
Compared with Waleffe's ODE model, the other qualitatively
new couplings introduced in~(\ref{eqn:a1})
- (\ref{eqn:a8}) are the term $\beta (A_3 A_6'')'$ in~(\ref{eqn:a5})
and the term $-2\alpha^2\beta A_1' A_6$ in~(\ref{eqn:a8}).

%%%%%%%%%%%%%%%%%%%%%%%%%%%%%%%%%%%%%%%%%%%%%%%%%%%%%%%%%%%%%%%%%%%%%%%%
\section{Numerical results}
\label{sec:results}

In this section we present the results of time-stepping
the system of PDEs~(\ref{eqn:a1}) - (\ref{eqn:a8}) over the range
$50 \leq Re \leq 200$. Our numerical method is the pseudospectral
exponential
time-stepping scheme referred to as `ETD2' by Cox \& Matthews (2002), using
128 Fourier modes in $z$, and suitably small timesteps such that
our results were insensitive to the timestep used.

As in Moehlis et al (2004) our primary interest is in the
emergence of a chaotic saddle in phase space. This is indicated by the 
lifetimes of chaotic transients as trajectories
evolve towards the linearly stable laminar state,
having started from initial conditions that are far from the
laminar equilibrium.

%%%%%%%%%%%%%%%%%%%%%%%%%%%%%%%%%%%%%%%%%%%%%%%%%%%%%%%%%%%%%%%%%%%%%%%%%%
\subsection{Initial conditions and domain parameters}

The lifetimes of chaotic transients are of course sensitive to the choice of 
initial condition, and in order to investigate the response of the
flow to a spatially localised
perturbation we took initial conditions corresponding to
a Gaussian profile, scaled so 
that the four modes $A_3$, $A_4$, $A_5$ and $A_6$
gave equal contributions to the initial kinetic energy $\KE_0$.
Later, in subsection~\ref{sec:results}(\ref{subsec:sin}),
we compare these results
with those obtained using a sinusoidal initial condition.

Specifically our Gaussian initial condition takes the
form $A_1=A_2=A_7=A_8=0$ and
\ba
A_j = c_j \exp\left( -\frac{(z-L_z/2)^2}{2\sigma^2} \right),
\label{eqn:gaussian}
\ea
for $j=3,\ldots,6$, where the coefficient $\sigma$ describes the
width of the Gaussian, and the normalisation constants $c_j$ are given by
\ba
c_3 & = & \sqrt{\frac{\KE_0\sigma\alpha}{2\pi \sqrt{\pi} \left(\beta^2\sigma^2
+ \frac{1}{2} \right) }}, \label{eqn:norm3} \\
c_4 = c_6 & = & \sqrt{\frac{\KE_0\sigma\alpha}{2\pi \sqrt{\pi}\left(\alpha^2\sigma^2
+ \frac{1}{2} \right)}}, \label{eqn:norm46} \\
c_5 & = & \sqrt{\frac{\KE_0\sigma\alpha}{\pi \sqrt{\pi}\left(\alpha^2\sigma^2
+ \frac{1}{2} \right)}}. \label{eqn:norm5}
\ea
The differencies in the expressions for $c_3,\ldots,c_6$ reflect
the different contributions made by $A_3,\ldots,A_6$ to the kinetic
energy~(\ref{eqn:etilde}).

We present results for a domain size $L_x=1.75\pi$, $L_z=1.2\pi$ corresponding
to the `minimal flow unit' identified by previous authors (Hamilton, Kim \&
Waleffe, 1995, and adopted by Moehlis et al 2004) as the smallest
domain in which sustained spatiotemporally chaotic dynamics have been
found numerically. We consider values for $\sigma$ in the range
$0.2 \leq \sigma \leq 0.8$, initially setting $\sigma=0.2$,
so that the initial Gaussian disturbance
is always spatially well-localised in the $z$ direction.

%%%%%%%%%%%%%%%%%%%%%%%%%%%%%%%%%%%%%%%%%%%%%%%%%%%%%%%%%%%%%%%%%%%%%%%%%%%%%%%%%%%%
\subsection{Results at fixed Reynolds numbers}

Our numerical integrations are carried out until either the
dynamics approaches very close to the laminar state, or
 until 1000 dimensionless time units of $h/(2U_0)$
have elapsed: this is the maximum lifetime
that transients are followed for in our computations. We follow transients
for the range $50 \leq Re \leq 200$, increasing $Re$ in steps of unity, and
increasing initial kinetic energy
$\KE_0$ in the range $0 < \KE_0<3.0$ in steps of $0.01$.

\begin{figure}[!h]
\bc
\includegraphics[width=13.0cm]{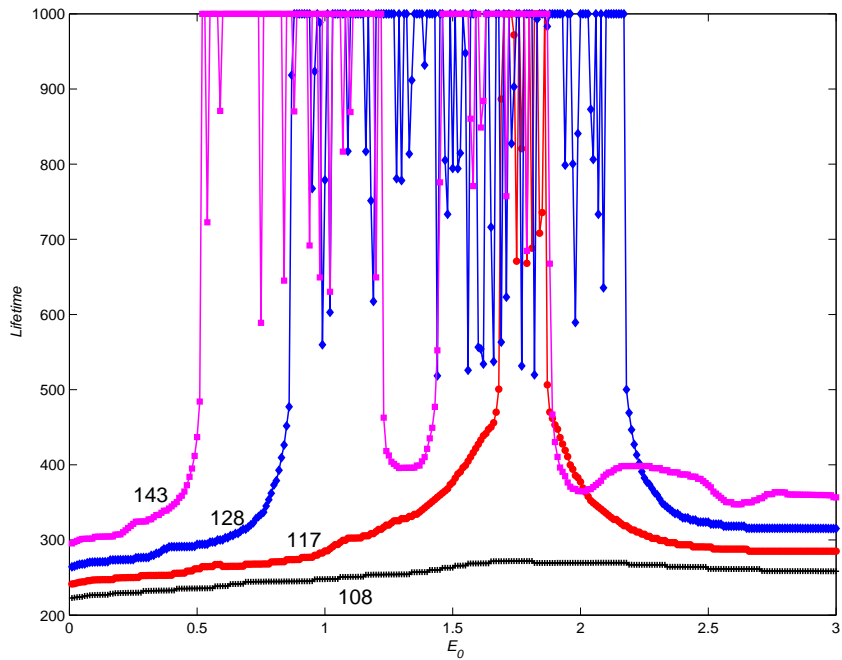}
\caption{(Online version in colour.)
Lifetimes of transients started from the Gaussian initial
condition, for initial kinetic energies $0<\KE_0<3.0$ for
$Re=108$ (black, $+$), $Re=117$ (red, $\circ$), $Re=128$ (blue, $\Diamond$) and
$Re=143$ (magenta, $\Box$).
Parameters are $\sigma=0.2$, $L_x=1.75\pi$, $L_z=1.2\pi$.}
\label{fig:lines}
\ec
\end{figure}

At low Reynolds numbers, $Re < 117$, we find numerically that
all initial conditions decay monotonically towards the
stable laminar equilibrium,
with a lifetime that increases slowly with $\KE_0$ in the range
$0<\KE_0< 1.7$ approximately, and then decreases slowly as $\KE_0$
increases further.
For $Re \geq 117$ the dynamics changes abruptly and initial energies around $\KE_0=1.7$
show far longer transients, as indicated in figure~\ref{fig:lines} which shows
the lifetimes of trajectories for four fixed values of $Re$, as $\KE_0$ varies.
The overall impression of figure~\ref{fig:lines} is of a well defined
transition from laminar to spatio-temporally complex dynamics for
a range of initial perturbations. It is clear that windows of rapid
attraction to the laminar state remain, for example for $Re=143$ and
initial energies around $\KE_0=1.3$.

%%%%%%%%%%%%%%%%%%%%%%%%%%%%%%%%%%%%%%%%%%%%%%%%%%%%%%%%%%%%%%%%%%%%%%%%%
\subsection{Transition boundary and structure of the chaotic saddle}

\begin{figure}[!h]
\bc
\includegraphics[width=13.0cm]{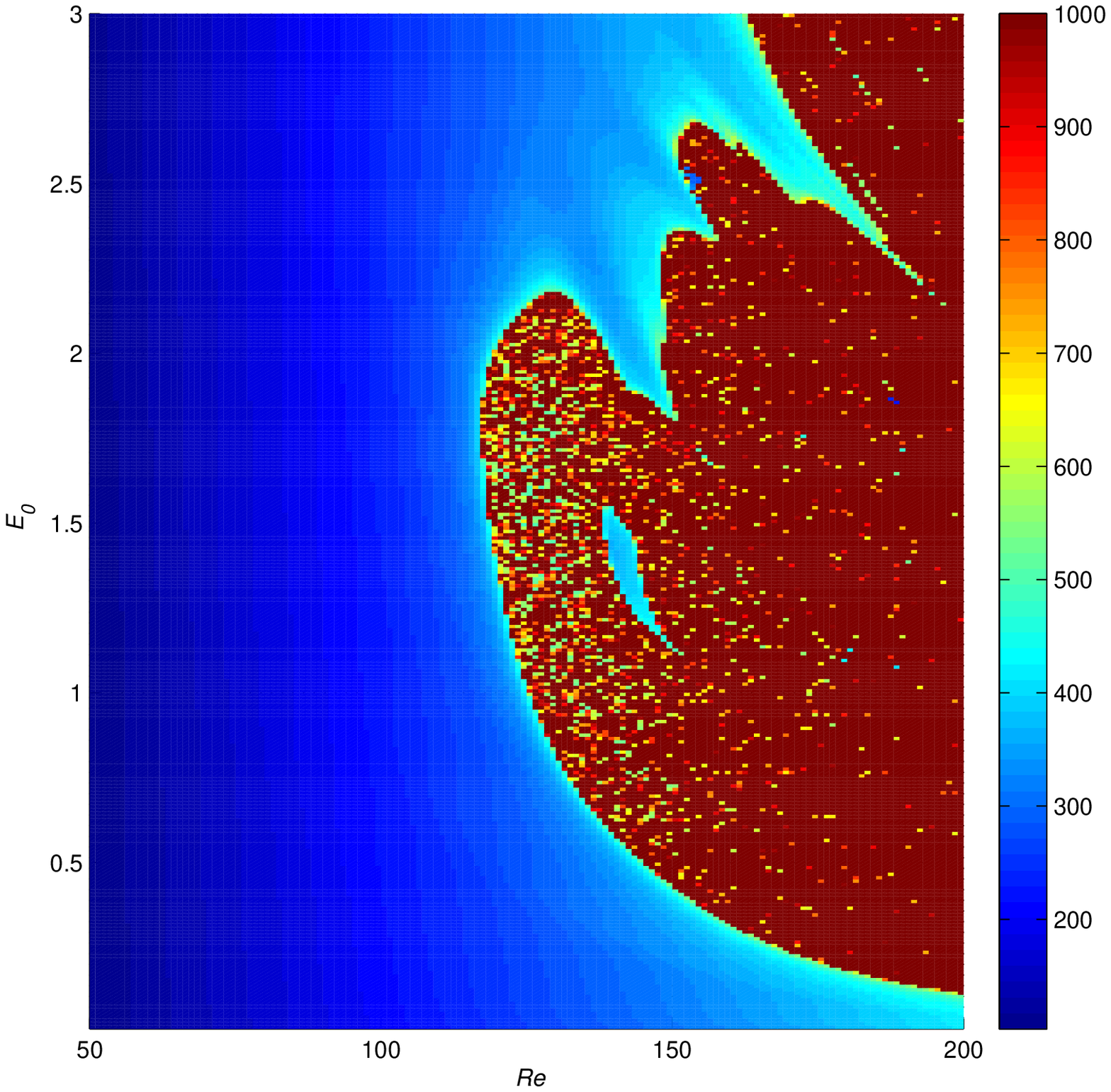}
\caption{Lifetimes of transients started from the Gaussian initial
condition, over the range $0 < \KE_0 < 3.0$ and
$50 \leq Re \leq 200$. Parameters are $\sigma=0.2$, $L_x=1.75\pi$, $L_z=1.2\pi$.}
\label{fig:sigma0.2}
\ec
\end{figure}

The well defined `transition boundary' at which there
is an abrupt increase in lifetimes
is shown clearly in figure~\ref{fig:sigma0.2} which presents a colour-coded
surface plot of lifetimes as $Re$ and $\KE_0$ vary. In contrast to
previous similar plots, for example figure~6 of Moehlis et al. (2004),
and see also figure~\ref{fig:sin} in the present paper, where
spatio-temporally complicated transients appear at around $Re=120$
at the ends of `wispy' dendritic fingers that coalesce as $Re$ increases,
figure~\ref{fig:sigma0.2} indicates the sudden appearance of a `fatter'
chaotic saddle in phase space.
The chaotic saddle appears to become less `porous' as $Re$ increases, and
there appears to be only one substantial hole, at around $Re=140$, $\KE_0=1.4$.
As in Moehlis et al. (2004) we might
anticipate that figure~\ref{fig:sigma0.2} has
fine scale structure as we zoom in. In contrast to the results
of that paper we find however that finer-scale investigations
do not reveal any coherent organisation to the lifetimes: instead
of regions of concentric bands of different colours, for example,
we find merely fine-scale apparent randomness and intermittency.
\begin{figure}[!h]
\bc
\includegraphics[width=6.0cm]{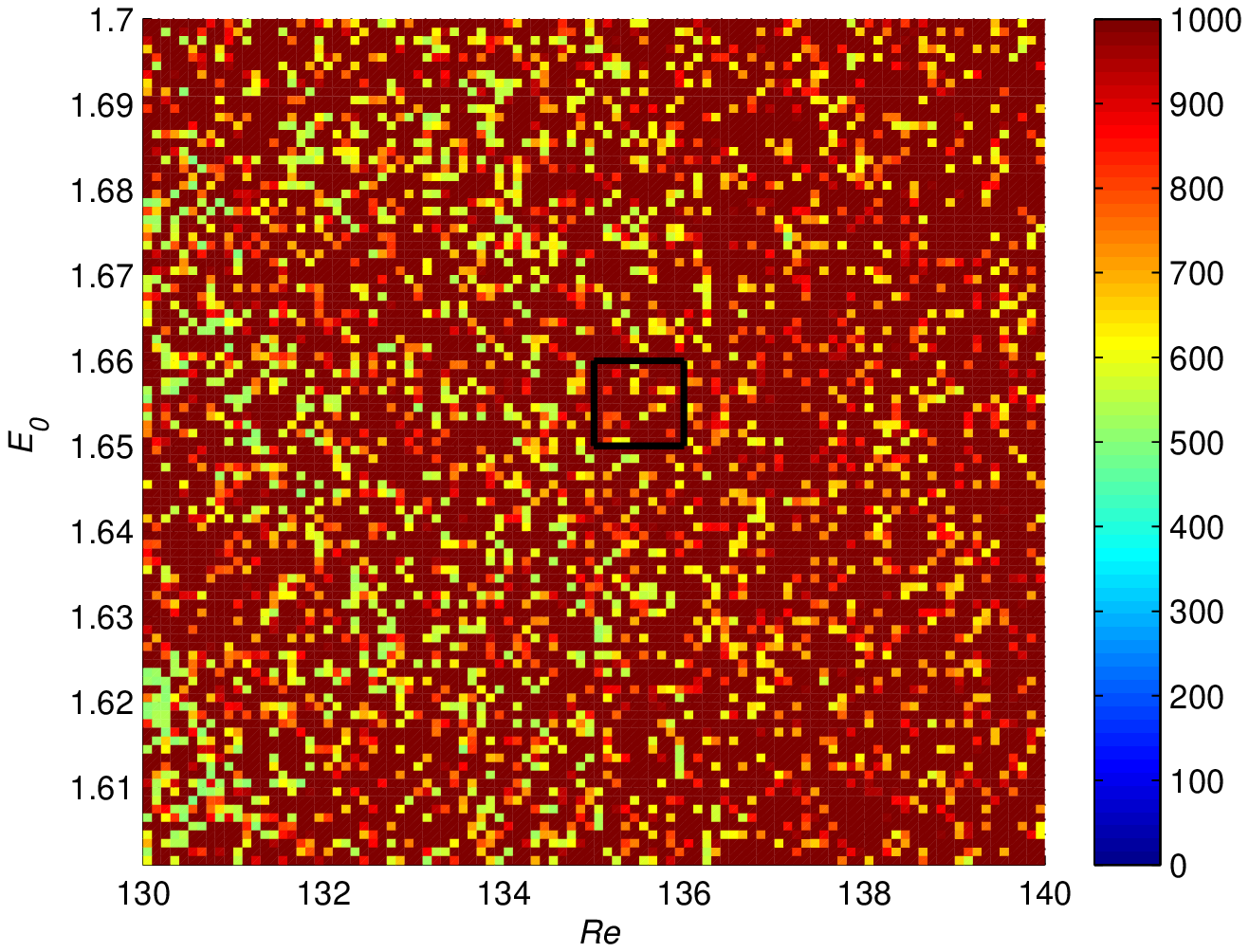}  % zoom-fig002.eps
\includegraphics[width=6.0cm]{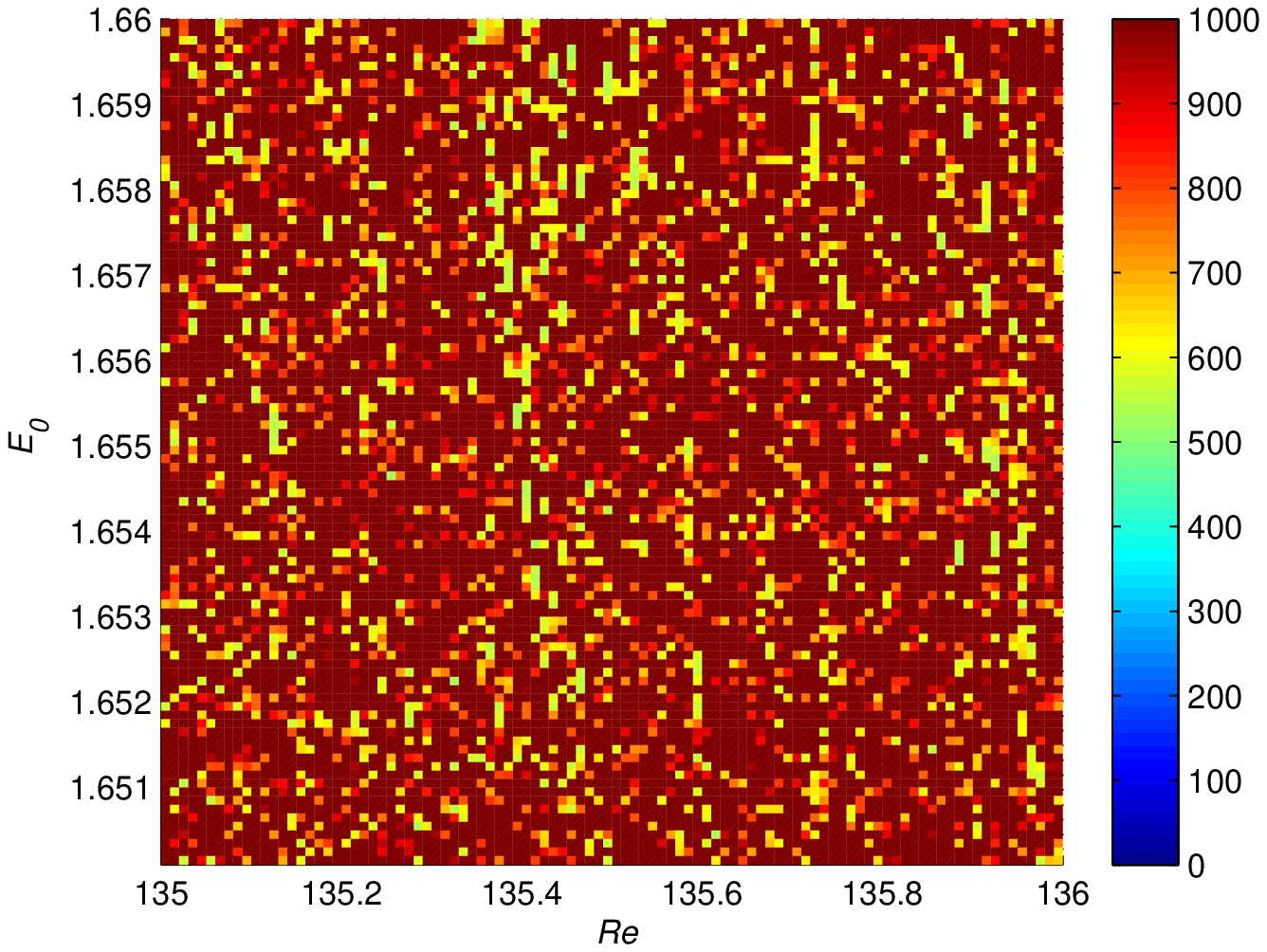}  % zoomzoom-fig002.eps
\centerline{(a) \hspace{5.75cm} (b)}
\caption{(Online version in colour.)
Enlargements of figure~\ref{fig:sigma0.2} showing a lack
of coherent organisation to the lifetimes at higher resolution in $Re$ and
$\KE_0$. (a) Transient lifetimes over the range $1.6< \KE_0 <1.7$ and
$130 \leq Re \leq 140$. (b) Transient lifetimes over the range
$1.65 < \KE_0 < 1.66$ and $135 \leq Re \leq 136$ as indicated by the
black square in (a). For both figures, the other
parameter values are $\sigma=0.2$, $L_x=1.75\pi$, $L_z=1.2\pi$.}
\label{fig:zoom}
\ec
\end{figure}
This is illustrated in
figure~\ref{fig:zoom}. Figure~\ref{fig:zoom}(a)
shows the region $130 \leq Re \leq 140$, $1.6 \leq \KE_0 \leq 1.7$
with a ten-fold increase in the resolution along each axis.
Figure~\ref{fig:zoom}(b) shows a further order of magnitude
increase in the
resolution both in $Re$ and in $\KE_0$; this part of the figure
corresponds to the region within the black box in the centre of
figure~\ref{fig:zoom}(a). This very detailed
enlargement appears to exhibit some slight
correlation, for example a propensity
to favour lifetimes of around 600 time units at $Re \approx 135.4$
but otherwise the unpredictable intermittent nature of the dynamics
continues as we consider finer and finer divisions in $Re$ and
$\KE_0$. This qualitative observation should be contrasted with
the lifetimes figures plotted by Moehlis et al. (2004) which
exhibit much more structure. We return to this point in
section~\ref{sec:results}(\ref{subsec:sin}).

\begin{figure}[!h]
\bc
\includegraphics[width=14.0cm]{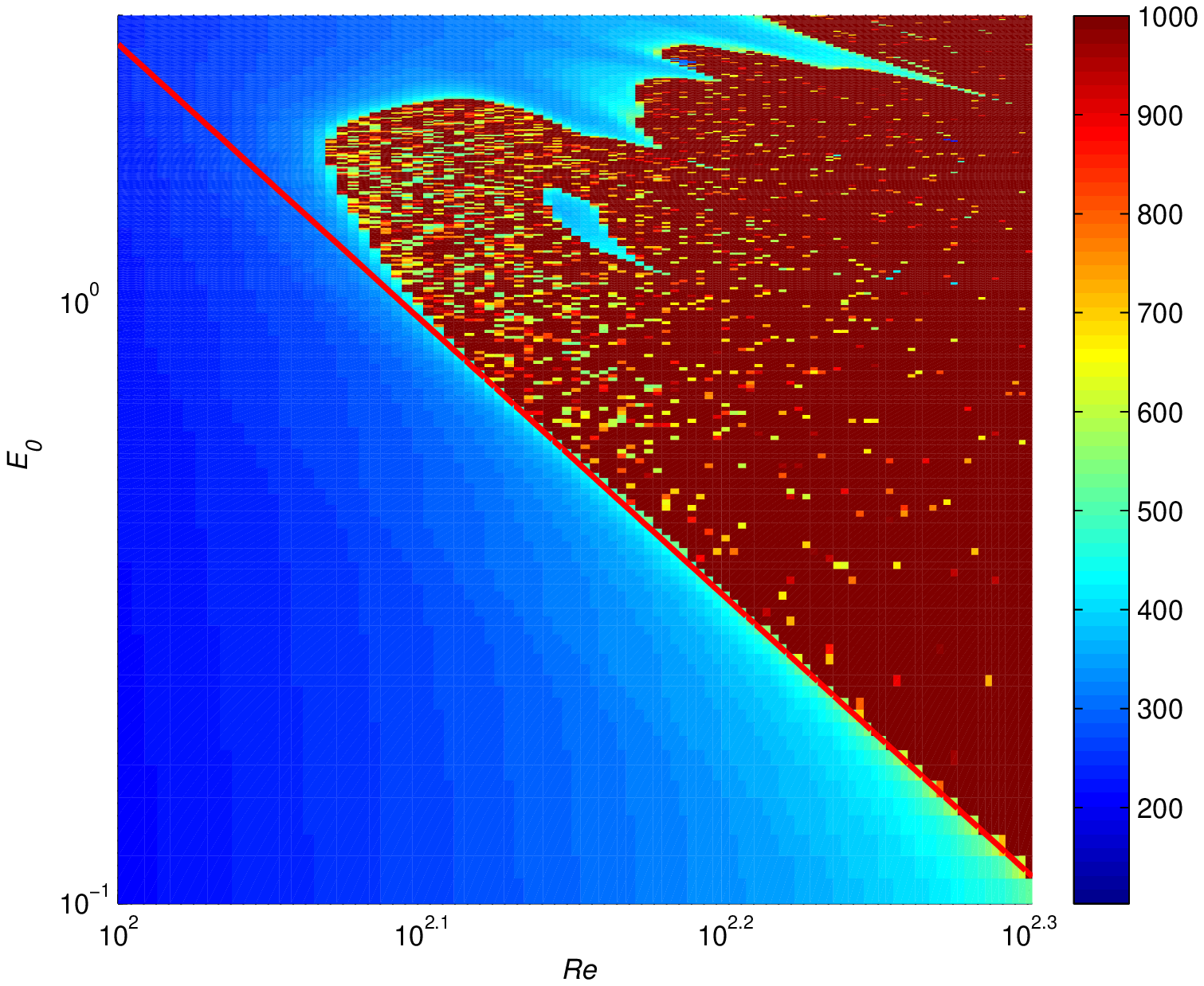} %all3-loglog001.eps
\caption{Log--log plot of transient lifetimes over the range $0.1<\KE<3.0$ and
$100 \leq Re \leq 200$ to show power law form of the lower boundary of the
attractor. The solid (red) line indicates the 
power-law $\KE_0 = \left( \frac{125}{Re} \right)^{4.65}$.
Parameter values are $\sigma=0.2$, $L_x=1.75\pi$, $L_z=1.2\pi$.}
\label{fig:sigma0.2loglog}
\ec
\end{figure}
\begin{figure}[!h]
\bc
\includegraphics[width=6.27cm]{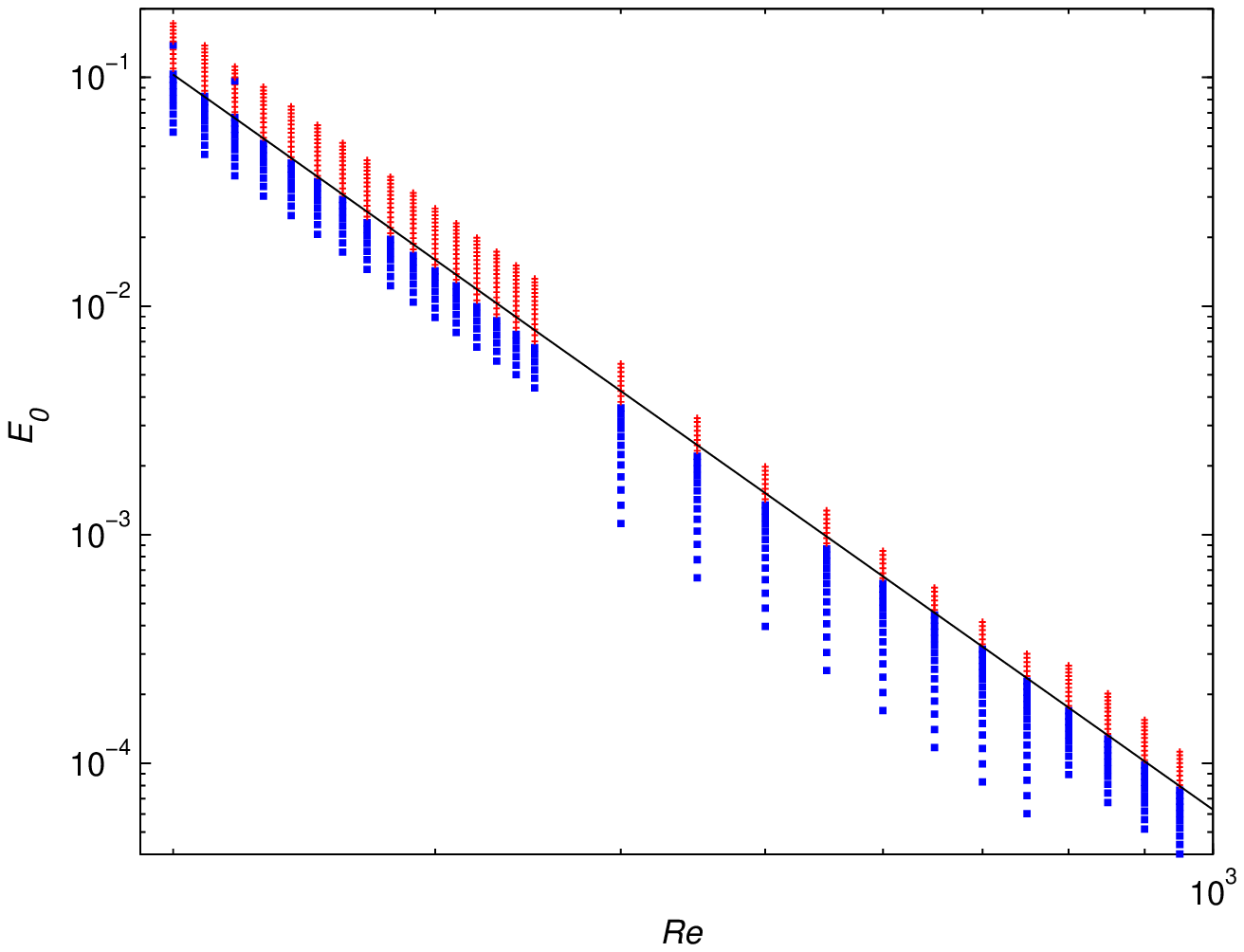} % bdry-to-1000.eps
\includegraphics[width=6.27cm]{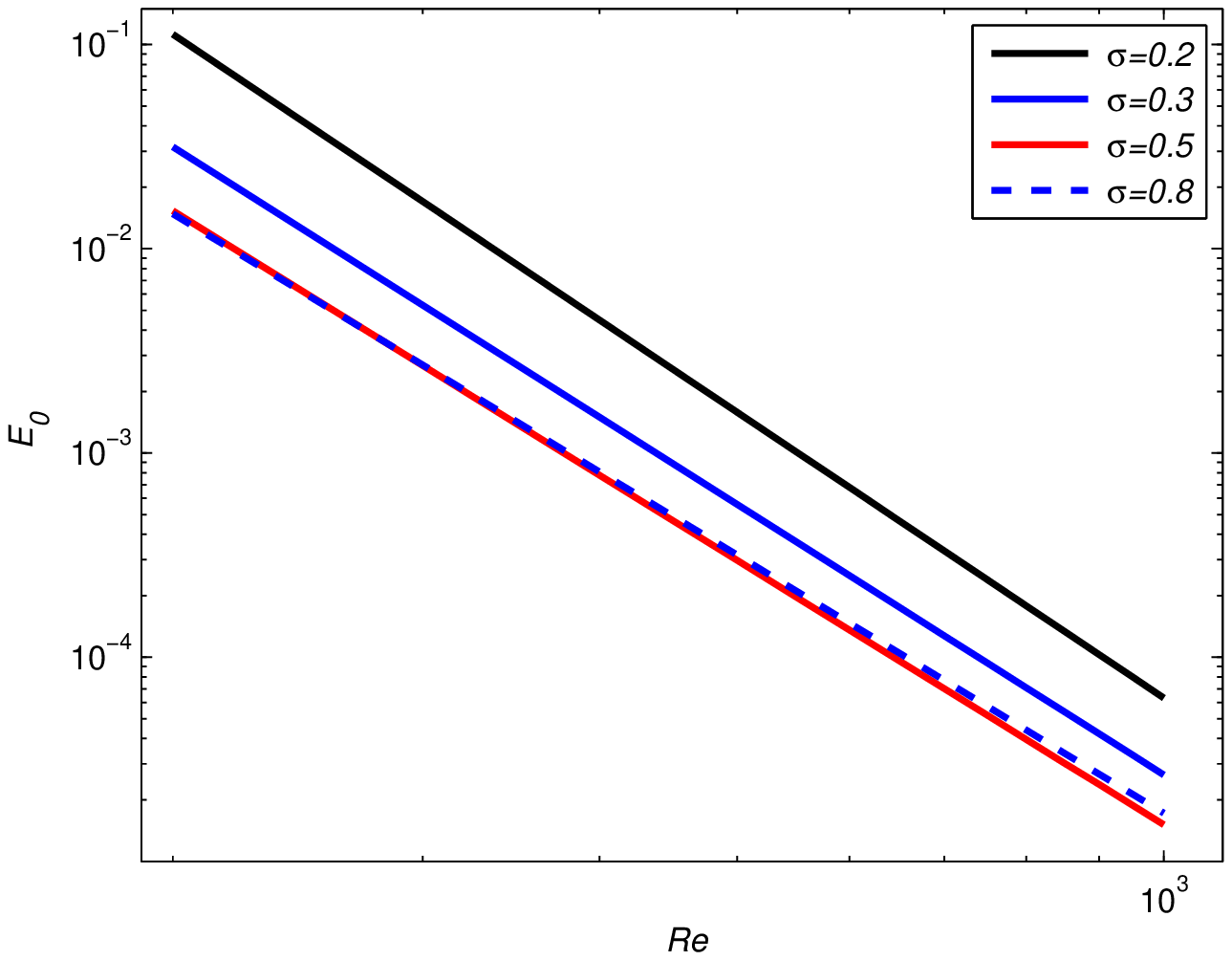} % varysigma002.eps
\centerline{(a) \hspace{6.00cm} (b)} 
\caption{(Online version in colour.)
Estimates of the lower boundary of the chaotic
saddle over the range $200< Re < 1000$. (a) $\sigma=0.2$;
squares (blue) indicate rapid return to the laminar state. `$+$' symbols
denote a long-lived chaotic transient.
The black line indicates the power-law scaling of the lower boundary.
(b) Best-fit power laws for a range of values of $\sigma$,
showing a broad insensitivity to the width parameter. 
Parameter values are $L_x=1.75\pi$, $L_z=1.2\pi$.}
\label{fig:higherRe}
\ec
\end{figure}
The lower boundary of the chaotic saddle is of particular physical interest,
since it describes the smallest amplitude perturbation required to produce
an extended chaotic transient. For the particular
form of perturbation used here, in the case $\sigma=0.2$,
we find numerically that the lower
boundary in figure~\ref{fig:sigma0.2} is remarkably well fitted by a
power law curve
%$\KE_0 = \left( \frac{125}{Re} \right)^{4.65}$
as illustrated in the log-log plot in figure~\ref{fig:sigma0.2loglog}.
Additional computations show that a power law continues to
provide a very good fit as $Re$ is increased,
up to at least $Re=10^3$, as shown in
figure~\ref{fig:higherRe}(a). The
black curve in figure~\ref{fig:higherRe}(a) is the best-fit curve
\ba
\KE_0 = \left( \frac{122}{Re} \right)^{4.6}. \label{eqn:powerlaw}
\ea
Blue squares in figure~\ref{fig:higherRe}
indicate that the flow returned to laminar before the computational
time limit $T_{\mathrm{max}}$ was reached, while red
$+$ signs indicate that the flow did not relaminarise before
$t=T_{\mathrm{max}}$. We set $T_{\mathrm{max}}=5000$ for $200<Re\leq 750$
and $T_{\mathrm{max}}=10000$ for $800 \leq Re \leq 950$.

\begin{table}
\caption{Dependence of power-law scalings on the width $\sigma$ of
initial condition.}
\bc
\begin{tabular}{ccc}
%\hline
%\hline
%\\
$\sigma$ & Constant, $c$  & Exponent, $p$ \\
\hline
$0.2$   &     $122$        & $4.6$ \\
$0.3$   &     $91.2$       & $4.4$ \\
$0.5$   &     $75.7$       & $4.3$ \\
$0.8$   &     $73.4$       & $4.2$
\\
\hline
\end{tabular}
\ec
\label{table:scalings}
\end{table}

In figure~\ref{fig:higherRe}(b) we show the best-fit power-law for
the boundary between laminar and spatio-temporally complicated flow
for different widths of the Gaussian initial
condition~(\ref{eqn:gaussian}). All are excellent fits to the data over
the range $200<Re<1000$, and the results are identical for computations
using 128 modes or
32 modes in $z$. Since the slopes of the power-laws become more negative
as the coefficient increases, the overall envelope of perturbation energies
as $Re$ increases, formed by taking the minimum value over all four curves,
has contributions from each curve at sufficiently large $Re$. This
envelope is an estimate of the true lower boundary for which we
would ideally compute the minimum over all possible forms of
initial perturbation.

For perturbations with a Gaussian profile~(\ref{eqn:gaussian}), and
for the range of $\sigma$ considered here,
it appears unlikely that the differences in exponents would
be detectable experimentally over the range
of relevant and accessible $Re$.
Table~\ref{table:scalings} gives the details of the best-fit parameters
of the power-laws shown in figure~\ref{fig:higherRe}(b), using
a functional form $E_0=(c/Re)^p$.

%%%%%%%%%%%%%%%%%%%%%%%%%%%%%%%%%%%%%%%%%%%%%%%%%%%%%%%%%%%%%%%%%%%%%%%%%%
\subsection{Lifetime distributions}

Moehlis et al also computed the distribution of lifetimes of turbulent
transients at fixed initial energies and Reynolds numbers.
They found that the survival probability
$P(T)$ that the solution had not decayed back to the laminar
state after a time $T$ was distributed exponentially, with
a mean lifetime that increased rapidly with $Re$ above the
critical value at which the chaotic saddle appeared. Numerically,
we construct a lifetime distribution by distributing the
initial energy randomly across a subset of the modes.
For our spatially-extended
model there are two possible approaches to
randomising the distribution of energy.

In the first approach, the total initial energy is
distributed uniformly on a spherical energy shell. This can be achieved
straightforwardly by modifying the expressions~(\ref{eqn:norm3})
- (\ref{eqn:norm5}) for the normalisation coefficients by
replacing $\KE_0$ with $4 \KE_0 \xi_j^2$ in the
expression for coefficient $c_j$, where $j=3,\ldots,6$. The coordinates
$\xi_j$ are those of points distributed at random over the unit sphere
in $\mathbb{R}^4$ given by $\sum_{j=3}^6 \xi_j^2 = 1$. We refer
to this as randomising over the amplitudes of the perturbations.

In the second approach, the central position of each
Gaussian perturbation $A_j$ is shifted randomly
in $z$ away from the centre of
the domain $L_z/2$, i.e. the values of the
coefficients $c_3,\ldots,c_6$ are
left unchanged but the centre of the Gaussian in~(\ref{eqn:gaussian})
is modified for each amplitude $A_j$. Since we employ
periodic boundary conditions, the total energy in the perturbation
is preserved. We refer to this procedure as randomising over
the locations of the perturbations.
\begin{figure}[!h]
\bc
\includegraphics[width=6.27cm]{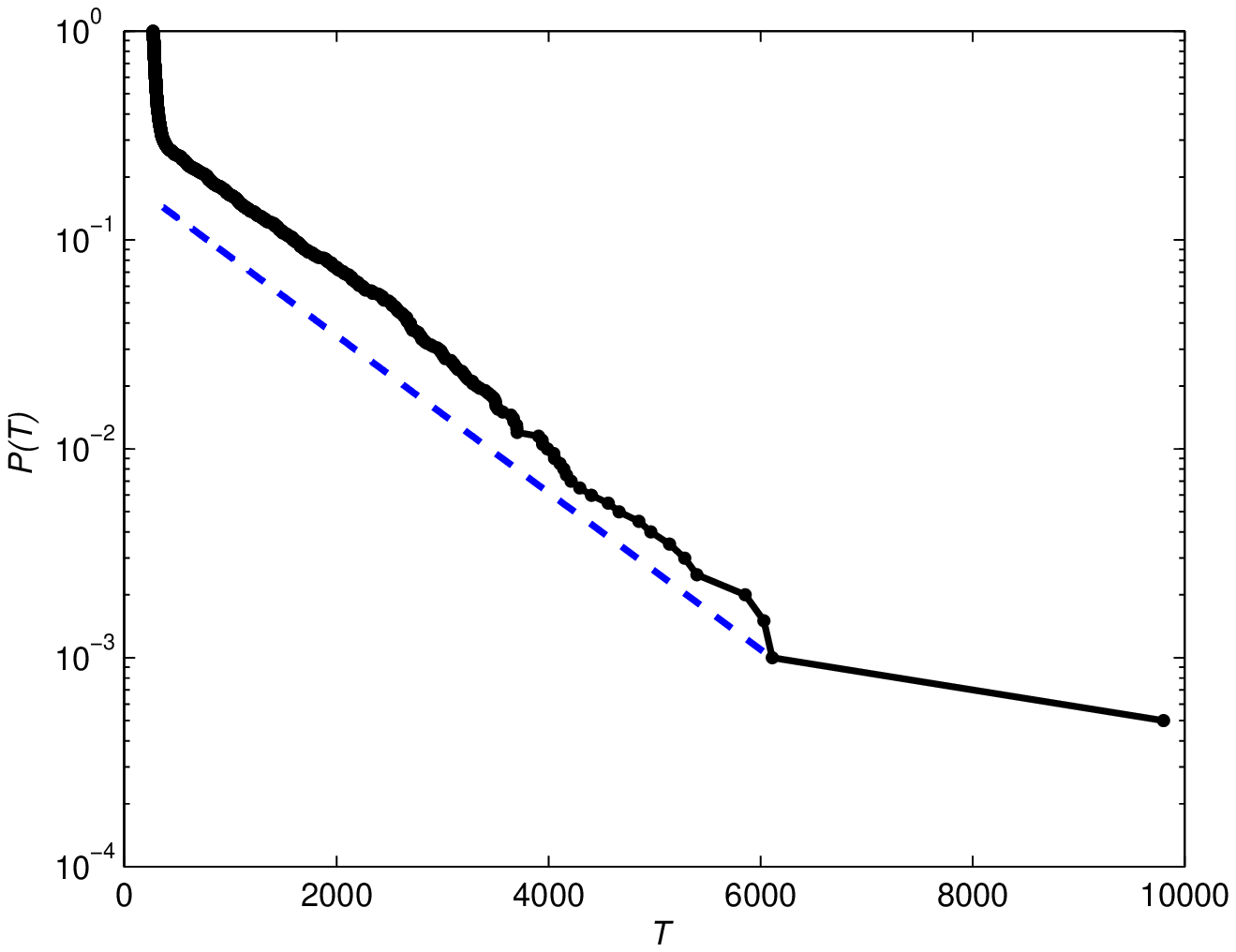} % {Re130-lifetimedist-004.eps}
\includegraphics[width=6.27cm]{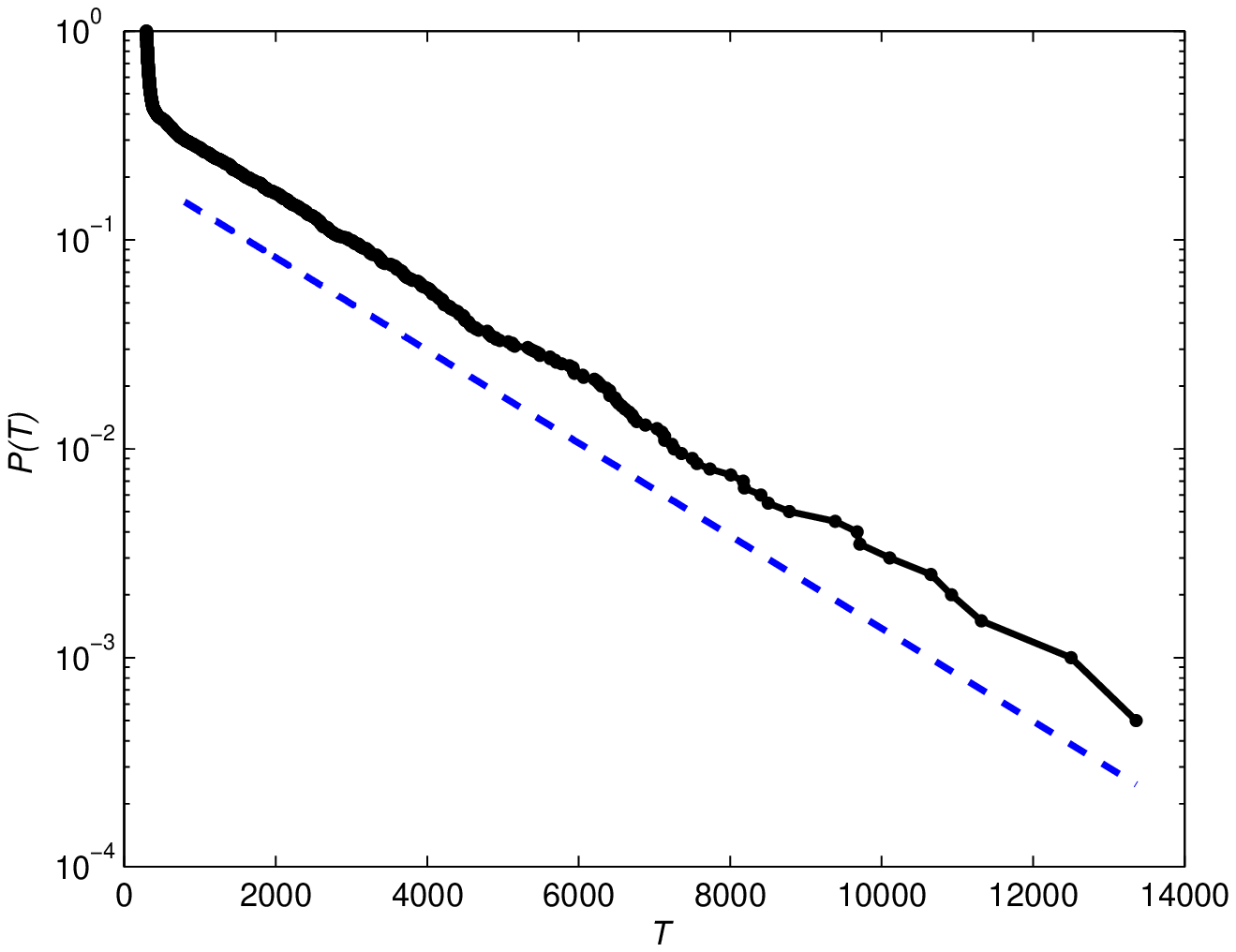} % {Re140-lifetimedist-004.eps}
\centerline{(a) \hspace{6.00cm} (b)} 
\caption{(Online version in colour.)
Distribution of lifetimes $P(T)$ computed over an ensemble of
2000 initial conditions produced by randomising over amplitudes
of perturbations at $\KE_0=1.0$. (a) $Re=130$ (b) $Re=140$.
Parameter values are $\sigma=0.2$, $L_x=1.75\pi$, $L_z=1.2\pi$.}
\label{fig:lifetimes-amp}
\ec
\end{figure}
Figure~\ref{fig:lifetimes-amp} shows the results of the first approach
at two Reynolds numbers quite close to the transition boundary. At
low lifetimes ($T < 200$) the survival probability remains unity since
any transient takes at least this finite amount of time
to decay to within the prescribed threshold of the laminar state. As $T$ increases there is an initial sharp drop
indicating that a substantial proportion of trajectories do decay
rapidly, with lifetimes in the range $200<T<400$. At larger $T$ the distribution
becomes close to a straight line on the linear-log plot, which is consistent
with an exponential distribution $P(T) \sim a_1 \exp(-a_2 T)$ for
$T \geq 400$ (approximately),
as indicated by the dashed line.
For $Re=130$ the best-fit coefficient values
are $a_1=3.93 \times 10^{-1}$ and $a_2=8.65\times 10^{-4}$. 
For $Re=140$ the best-fit coefficient values are $a_1=4.57 \times 10^{-1}$,
$a_2=5.11 \times 10^{-4}$. 
\begin{figure}[!h]
\bc
\includegraphics[width=7.0cm]{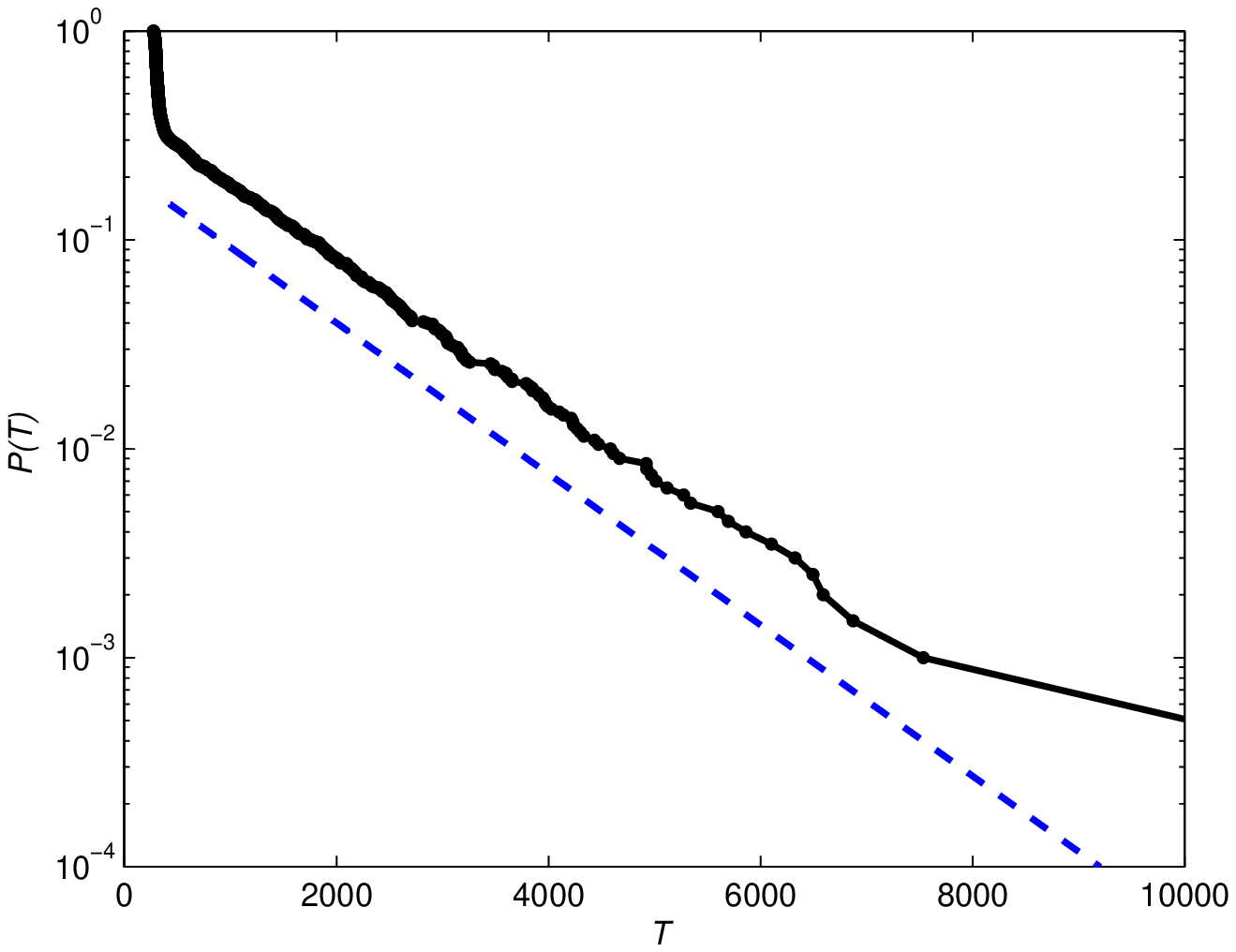} % {Re130-space-lifetimedist-001.eps}
\caption{(Online version in colour.)
Distribution of lifetimes $P(T)$ computed over an ensemble of
2000 initial conditions produced by randomising over the locations
of perturbations at $\KE_0=1.0$ and $Re=130$.
Parameter values are $\sigma=0.2$, $L_x=1.75\pi$, $L_z=1.2\pi$.}
\label{fig:lifetimes-loc}
\ec
\end{figure}
Similarly, figure~\ref{fig:lifetimes-loc} shows the lifetime
distribution computed from the second approach, randomising over
the locations of the perturbations. The distribution shows very
similar characteristics to that in figure~\ref{fig:lifetimes-amp}(a)
and is again well-described by an exponential distribution with
best-fit parameters $a_1=4.24 \times 10^{-1}$ and $a_2=8.32 \times 10^{-4}$.
We observe that the values for the coefficient $a_2$ are very similar
between the two randomisation methods.
In all cases, we expect that as $Re$ increases the
distribution exhibits systematic
deviation from exponential and flattens out. We anticipate that,
as in the ODE case, this is due to the appearance
of stable attractors near the chaotic saddle, as $Re$ increases.
These attracting sets then absorb trajectories
with positive probability, leading to
a proportion of trajectories which never return to the
laminar state (Moehlis et al 2005).

%%%%%%%%%%%%%%%%%%%%%%%%%%%%%%%%%%%%%%%%%%%%%%%%%%%%%%%%%%%%%%%%%%%%%%%
\subsection{Spanwise resolution}

The very high resolution used in the $z$ direction ($N=128$ Fourier modes)
is greater than required,
for such a small domain,
in order to capture the dynamics of the `active' modes of the system.
In order to probe the range of wavenumbers that make substantial
contributions to the dynamics we investigated the effect of
varying the truncation level of the numerical scheme in $z$. We
define the `$m$-mode truncation' of the PDEs~(\ref{eqn:a1}) - (\ref{eqn:a8})
by keeping the Fourier modes $\sim \e^{\pm \i n \gamma z}$
for $0 \leq n \leq m-2$
for $A_1$, $A_4$ and $A_5$, and keeping the modes for $0 \leq n \leq m-1$
for $A_2$, $A_3$, $A_6$, $A_7$ and $A_8$. The $m$-mode
truncation is thus a collection of (at most)
$16m-14$ real ODEs. This is an upper bound on the effective
dimension of the ODE dynamics since not every real and imaginary
part is coupled for every $m$.

\begin{figure}[!h]
\bc
\includegraphics[width=6.27cm]{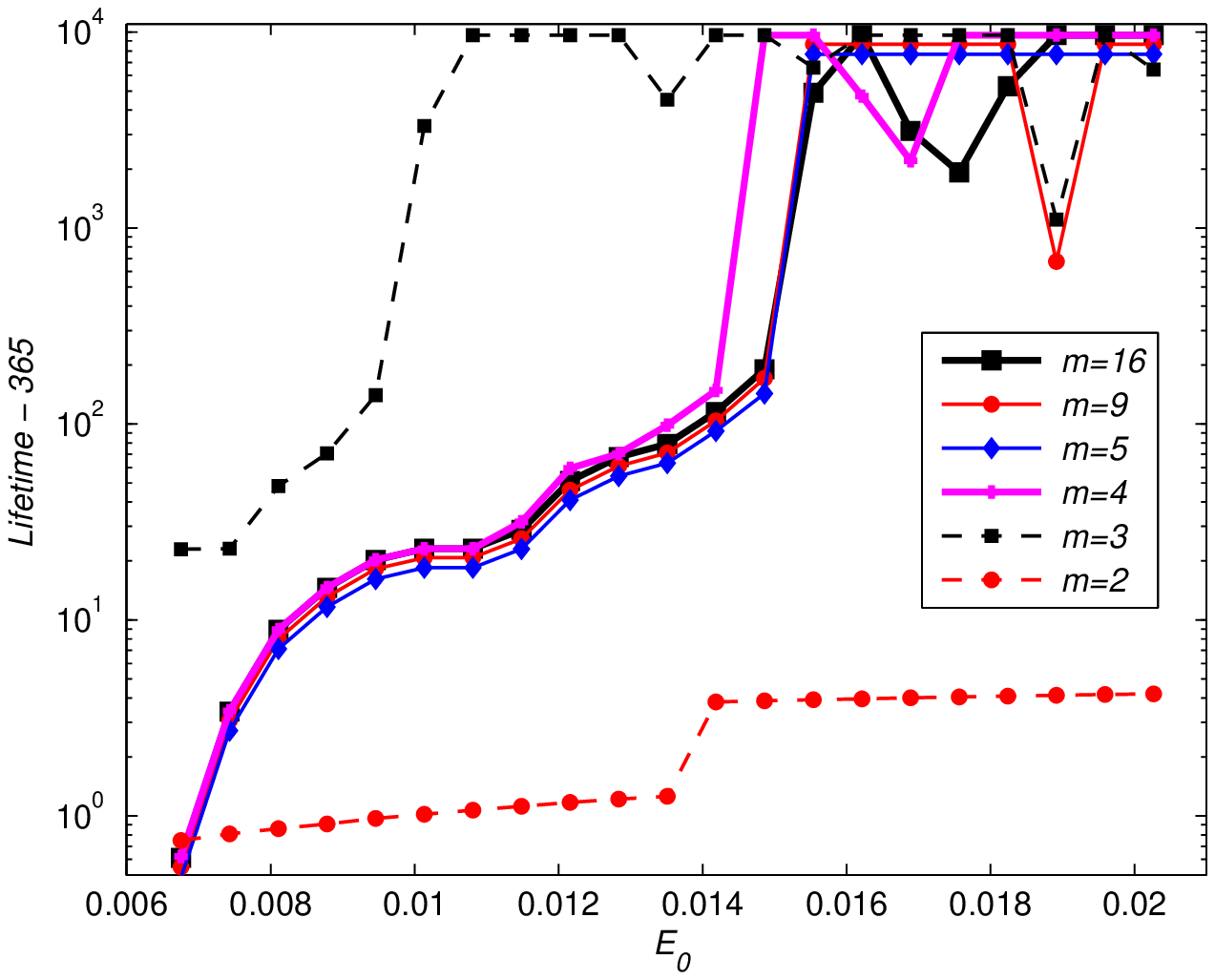} % {R200-varymm.eps}  %% was width=6.0cm
\includegraphics[width=6.27cm]{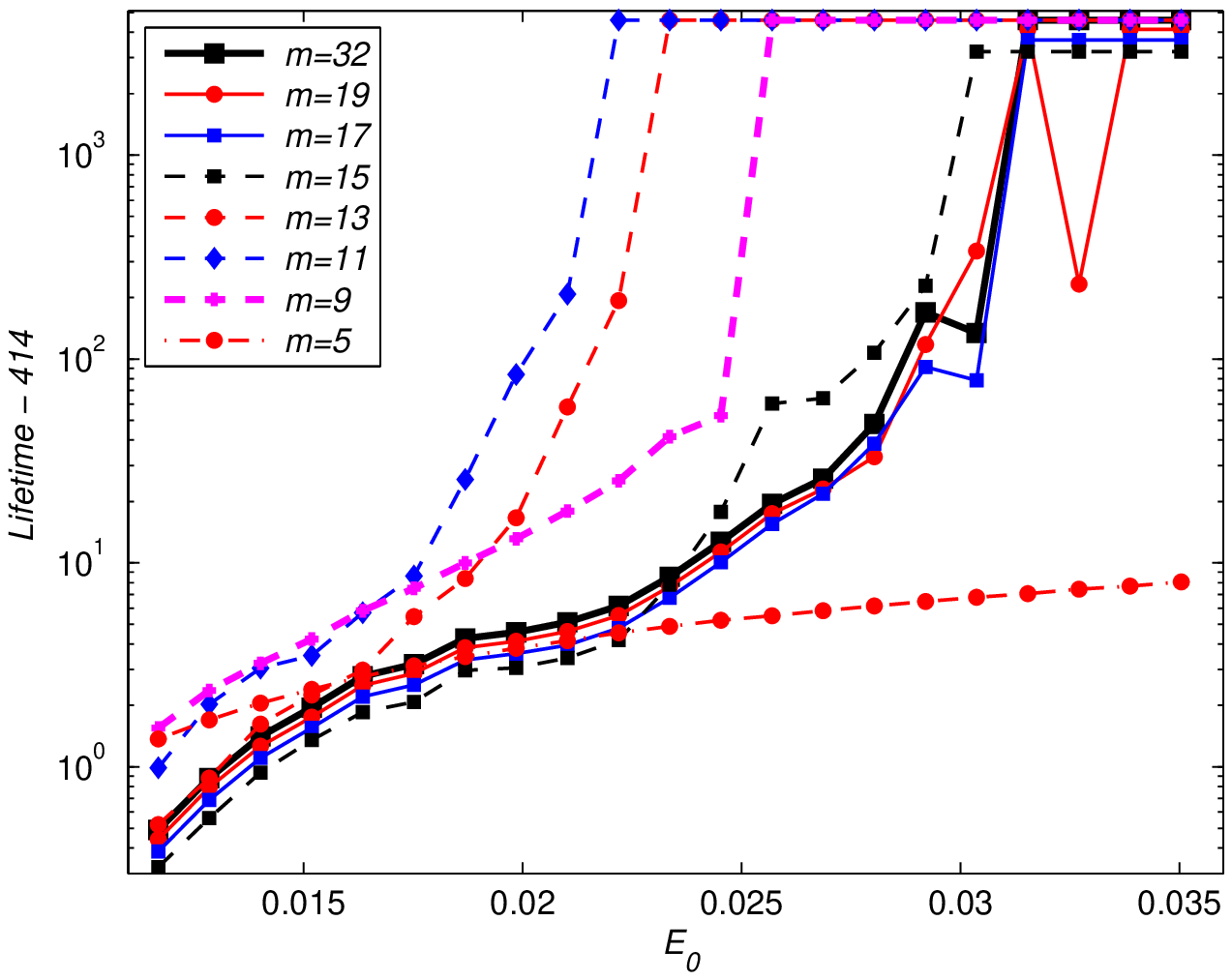} % {R200-varymmLzx5.eps}
\centerline{(a) \hspace{6.00cm} (b)} 
\caption{(Online version in colour.)
Lifetimes of transients $T$ as a function of initial
energy $E_0$ for $Re=200$ in domains of widths (a) $L_z=1.2\pi$
and (b) $L_z=6\pi$, showing the effect of varying the
truncation level $m$. Vertical axis is logarithmic for clarity,
and shows the additional lifetime after subtracting a constant:
$T-365$ in (a), $T-414$ in (b). In (a) the data for $m=9$
and $m=5$ lie exactly on the $m=16$ values: they are artificially
offset vertically for clarity. In (b) the data for $m=19$,
$m=17$ and $m=15$ lie exactly on the $m=32$ values and are
similarly offset for clarity.
Computations were terminated at $T_{\mathrm{max}}=5000$.
Parameter values are $\sigma=0.8$, $L_x=1.75\pi$.}
\label{fig:trunc}
\ec
\end{figure}

By construction, this further reduction
resembles the Waleffe model in the particular case $m=2$,
but by varying $m$ we are able to
probe the influence of the higher-wavenumber modes
on the location of the lower
boundary for the onset of temporally complex dynamics.
For each value of $E_0$ and $m$, a single initial condition
was used. This initial condition was the projection
of the Gaussian profiles
described in section~\ref{sec:results}(a) onto the available
Fourier modes in $z$. Results are shown in figure~\ref{fig:trunc}.
Figure~\ref{fig:trunc}(a), for the small domain $L_z=1.2\pi$,
 shows that computations with
truncation levels $m \geq 5$ produce an identical indication
of the boundary between laminar and spatio-temporally complex
behaviour. The case $m=4$ is qualitatively but not quantitatively
correct. For the larger domain $L_z=6\pi$, figure~\ref{fig:trunc}(b)
indicates that the numerical results are essentially
unchanged for $m \geq 17$. The number of Fourier modes
required to maintain accuracy in the larger domain is therefore
broadly in line with, although slightly lower than, what one might
naively expect.

Finally we note that as $m$ decreases further, the
boundary appears consistently to move to lower $E_0$. This behaviour
is purely a function of the organisation of invariant sets in
phase space; it is not clear that there is a
physical reason why these should appear to move in one direction or
the other.

%%%%%%%%%%%%%%%%%%%%%%%%%%%%%%%%%%%%%%%%%%%%%%%%%%%%%%%%%%%%%%%%%%%%%%%%%
\subsection{Spatio-temporal dynamics and effect of initial conditions}
\label{subsec:sin}

\begin{figure}[!h]
\bc
\includegraphics[width=10.0cm]{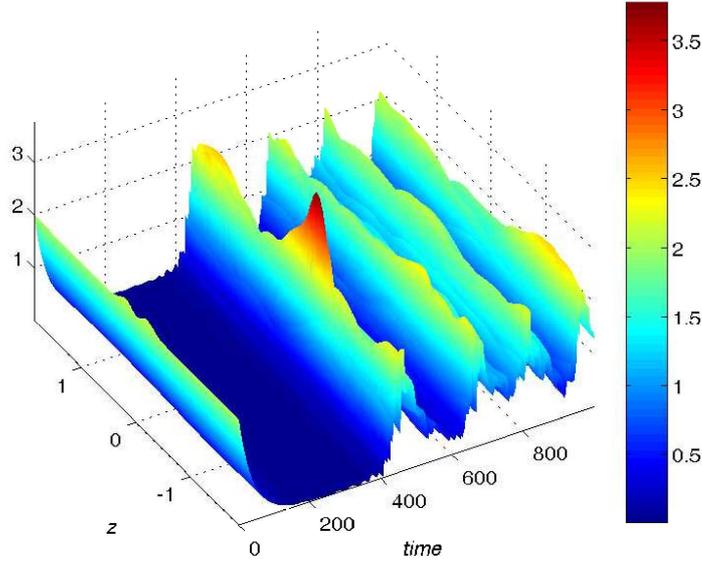} % {R200_Tinit0_11-persp002.eps}
\caption{(Online version in colour.)
Space-time plot of the local energy quantity
$\tilde{E}(z,t)-A_1^2+(A_1-\sqrt{2})^2$, 
indicated by both surface height and colour, for the
PDEs~(\ref{eqn:a1}) - (\ref{eqn:a8}) for $Re=200$
in the minimal flow unit domain. 
$\tilde{E}(z,t)$ is defined in equation~(\ref{eqn:etilde}).
The additional square terms
shift the laminar state to the zero level in the plot and
serve to highlight the localised `edge' state at $t\approx 450$.
Parameter values are $\KE_0=0.11$, $\sigma=0.2$, $L_x=1.75\pi$, $L_z=1.2\pi$.}
\label{fig:Re200}
\ec
\end{figure}

In this final subsection we comment on the spatio-temporal dynamics
of the PDEs for initial energies near
the transition boundary, and we compare the results
of section~\ref{sec:results}(c) for the lifetimes of transients
with those in this section obtained using sinsoidal initial conditions.
Figure~\ref{fig:Re200} shows the spatial and temporal evolution
of the PDEs for $Re=200$ and $\KE_0=0.11$, just above the transition
boundary. The quantity $\tilde{E}(z,t)-A_1^2+(A_1-\sqrt{2})^2$ is
plotted in the figure, so that the
laminar state $A_1=\sqrt{2}$ is at level zero, where
$\tilde{E}(z,t)$ is defined in~(\ref{eqn:etilde}).
The initial monotonic decay towards the laminar
state is interrupted by the growth of oscillatory disturbances.
These disturbances generate a sharp spike in the kinetic
energy, localised both in space and time, before the solution
settles into a spatio-temporally
chaotic state. For comparison, at $\KE_0=0.1$ we observe
only monotonic decay to the laminar state.

Analysis of the energy distribution across Fourier
modes shows that the small-scale oscillations just prior
to the localised spike
involve many higher-wavenumber Fourier modes. These mode amplitudes
grow very rapidly as we approach the 
spike state. Since the formation of the spike is, in almost every case,
the `edge state' that is the precursor
to spatio-temporally complicated dynamics,
one interpretation of the results in the
previous sub-section is that the spatial resolution required
to correctly determine the boundary of the basin of attraction
of the laminar state is indicated by the resolution required
to describe accurately these spatially-localised spikes.

More generally, the dynamics near the boundary between
monotonic relaminarisation and spatio-temporal complexity appear
to depend on the evolution of high-wavenumber modes, seeded by the
use of a Gaussian initial condition which injects energy into every
available mode. This is in contrast with the sinusoidal initial
conditions used in previous reduced models, where, by construction,
such a sinusoidal initial condition was the only possible choice.
\begin{figure}[!h]
\bc
\includegraphics[width=11.0cm]{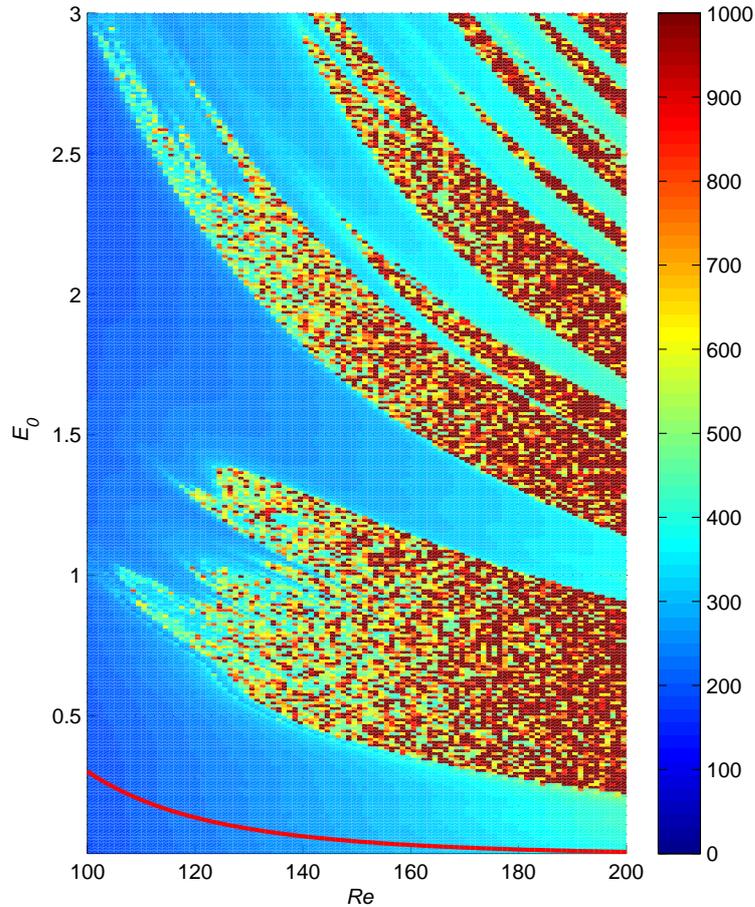} % {sin_ic_mm32-all.eps}
\caption{Lifetimes of transients started from the sinusoidal
initial condition, over the range
$0 < \KE_0 < 3.0$ and $100 \leq Re \leq 200$.
Solid (red) line indicates the boundary $\KE_0=(75.7/Re)^{4.3}$
corresponding to the lowest of the boundaries
in figure~\ref{fig:higherRe}(b) for Gaussian perturbations.
Parameter values are $L_x=1.75\pi$, $L_z=1.2\pi$. Computations
used $N=32$ Fourier modes in $z$ and were terminated
at $T_{\mathrm{max}}=1000$.}
\label{fig:sin}
\ec
\end{figure}

In figure~\ref{fig:sin} we show the analogous plot to
figure~\ref{fig:sigma0.2} for the lifetimes of transients,
but in this case using a sinusoidal initial condition
instead of a Gaussian profile.
For the sinusoidal initial condition we set
$A_2=c_2 \cos \gamma z$; $A_3=c_3 \sin \gamma z$; $A_4=c_4$; $A_5=c_5$;
$A_1=A_6=A_7=A_8=0$
where the constants $c_2, \ldots, c_5$ are given by
\ba
c_2=\sqrt{\frac{\alpha \KE_0}{2\pi L_z}}, \quad
c_3=\sqrt{\frac{\alpha \KE_0}{(\beta^2+\gamma^2)\pi L_z}}, \quad
c_4=\sqrt{\frac{\KE_0}{2\pi \alpha L_z}}, \quad
c_5=\sqrt{\frac{\KE_0}{\alpha \pi L_z}}, \nn
\ea
so that the inital kinetic energy $\KE_0$,
defined in~(\ref{eqn:etilde}),
is distributed equally between the four non-zero
amplitudes.

Comparing figures~\ref{fig:sigma0.2} and~\ref{fig:sin} there are
important qualitative and quantitative differences.
Firstly, the boundary
between relaminarisation and spatio-temporal complexity is much
more obvious in figure~\ref{fig:sigma0.2} where we
employ the Gaussian initial
condition. The boundary in figure~\ref{fig:sin} shows
much more `structure'; it is much less obvious that
a boundary in the sense, for example, of a countable
collection of (piecewise) continuous curves in the $(Re,\KE_0)$
plane, can even be defined.
Many deep valleys of short lifetimes
persist up to $Re=200$ and beyond.
Secondly, the solid (red) line in figure~\ref{fig:sin} shows the
lowest boundary computed for Gaussian perturbations, corresponding
to $\sigma=0.5$, see table~\ref{table:scalings}. It appears that
that the laminar state is substantially
more sensitive to Gaussian perturbations than those of
the same energy but in the form of 
low-wavenumber sinusoids. Tentatively, based only on the data in
figure~\ref{fig:sin}, we suggest that the
lower boundary of the appearance of spatiotemporally complex dynamics
arising from the sinusoidal perturbation scales as $\KE_0 \sim Re^{-2}$,
i.e. perturbation amplitude scaling as $Re^{-1}$.
This is a typical exponent produced by very many reduced models of
very low order, as summarised and discussed by Baggett \& Trefethen
(1997).

Within the region of spatio-temporally
complex dynamics, the lifetimes in figure~\ref{fig:sin} and in
additional enlargements (not shown here) show 
a degree of correlation between lifetimes at neighbouring points in the
$(Re,\KE_0)$ plane which is not nearly so clearly shown to
exist in figure~\ref{fig:sigma0.2}
or in the enlargements shown in figure~\ref{fig:zoom}.

In summary, in these respects
figure~\ref{fig:sin} is reminiscent of lifetime plots for
the very low-dimensional models of Moehlis et al (2004), see
their figures 6 and 8, and figures 5 and 6 in
the paper by Eckhardt \& Mersmann (1999).
We conclude that allowing for the accurate
representation of spatially-localised initial conditions
by extending the spanwise resolution of the model generates
results that differ substantially, both qualitatively and
quantitatively, from those of previous reduced models.

%%%%%%%%%%%%%%%%%%%%%%%%%%%%%%%%%%%%%%%%%%%%%%%%%%%%%%%%%%%%%%%%%%%%%%%
\section{Discussion and conclusions}
\label{sec:conc}

In this paper we have presented an extended version of the
Galerkin-truncated model due to Waleffe (1997) for the transition
to turbulence (or, at least, spatio-temporally complex dynamics)
in sinusoidal shear flow. This model is appealing since it provides
an intermediate step between previous analytical work and DNS of the full
Navier--Stokes equations. Preserving full resolution in the
spanwise ($z$) direction and removing the assumption of periodicity
allows both the use of spatially-localised initial conditions, and
the (transient)
formation of localised structures in the flow which (although unstable)
are known to exist and play a role in mediating the onset of
turbulence. The use of a small number of Fourier modes in the wall-normal
($y$) and streamwise ($x$) directions provides the simplification
of the underlying Navier--Stokes
equations, which in turn allows us to perform very detailed investigations
of the dynamics of this reduced model. 

We compare our results with those of previous authors,
in order to see which properties are common to
these different approaches, and which are not.
For example, we find, in agreement with the results of
Moehlis et al 2004, that the lifetimes of turbulent transients
are well-described by an exponential distribution.
However, our results show that the transition boundary,
while exhibiting some of the `structured' shape
observed by many authors (including, in the case
of pipe flow, Schneider, Eckhardt \& Yorke 2007), appears
at lower perturbation energies, and much more abruptly,
than for the ODE models investigated by
Eckhardt \& Mersmann (1999) and Moehlis
et al (2004). 
The PDE model that we present here is able to represent both
spatially-localised and spatially-extended initial conditions
and therefore we are able to make direct comparisons of this kind.

Our key finding is that spatially-localised initial conditions
are able to provoke complicated behaviour at substantially
lower energies than the sinusoidal, spatially-extended perturbations
used in previous studies. Moreover, the perturbation energy at the
lower boundary of the chaotic saddle
appears to scale as $Re^{p}$ with the exponent $p \approx -4.3$, rather
than $Re^{-2}$ as in Eckhardt \& Mersmann's 19-mode truncated ODE model
(note that their figure 5 showing an $Re^{-1}$ power law plots
$Re$ against mode amplitude which is proportional to $\KE_0^{1/2}$).
In the present work, the exponent in this power-law scaling was found
to depend only weakly on the width of the Gaussian perturbation used.

In addition, our results are robust to the numerical resolution used in
the spanwise direction, and, for a relatively small domain
of width $L_z=1.2\pi$, point to the necessity of
keeping around $5$ Fourier modes in $z$ in order accurately
to capture the dynamics of the fully-resolved PDE model.
One possible explanation of these results is that
admitting higher-wavenumber modes
generates many more invariant sets within the boundary
of the basin of attraction of the laminar state. Then, even small
amounts of initial energy in these modes forces the system
to spend much longer in the vicinity of these sets before being
able to relaminarise. In this sense, `holes' in the basin boundary
are filled in. The existence of these new invariant sets, and the
lack of `holes', leads to a robustness in the lengths of transients,
and therefore to a more clearly defined boundary between monotonic
relaminarisation and longer-lived transients.

It would clearly be of interest in future work to
look at the relation between
localised states which have been observed and studied in some
detail in DNS for shear flow problems (Schneider et al 2010a, 2010b)
and the dynamics of the reduced model presented here. We anticipate
that the reduced model contains such states, and the
homoclinic snaking bifurcation diagrams that typically organise them
in driven dissipative systems such as shear flows, just as
model ODE truncations, for example that discussed in
Moehlis et al 2005, contain equilibria and time-periodic solutions very
similar to those located in DNS (Nagata 1990; Gibson et al 2009).
It should be possible systematically to further
reduce the model equations presented
here in order to make direct connections between theoretical work
on localised states (Burke \& Knobloch 2006; Chapman \& Kozyreff 2009;
Dawes 2010)
and the DNS results referred to above. In turn, the identification
and analysis
of additional unstable invariant sets within the boundary of the basin
of attraction of the laminar state (as discussed by Lebovitz 2009),
and their parameter dependence, will greatly help our understanding
of the process of relaminarisation.

%%%%%%%%%%%%%%%%%%%%%%%%%%%%%%%%%%%%%%%%%%%%%%%%%%%%%%%%%%%%%%%%%%%%%%
\begin{acknowledgements}
JHPD would like to thank Rich Kerswell and
Tobias Schneider for useful conversations, and
Matthew Chantry for a minor correction.
Both authors are grateful to the anonymous referees for
very useful comments, and they gratefully acknowledge financial
support from the Royal Society; JHPD currently holds
a Royal Society University Research Fellowship.
\end{acknowledgements}

%Begin the bibliography here

\end{document}